\documentclass[useAMS,usenatbib]{mn2e} \usepackage{amssymb,graphicx,amsmath}

\def\hMsol{\, h^{-1}{\rm M_{\odot}}} 
\def\hMpc{\, h^{-1}{\rm Mpc}} 
\def\hkpc{\, h^{-1}{\rm kpc}} 
\def\kms{\, \hbox{km}\,\hbox{s}^{-1}}

\def\erghMsol{\, h\, {\rm erg}\, M_{\odot}^{-1}} 
 
\def\Gyr{\, {\rm Gyr}}

\def\rgb{}
\def\ahot{{\alpha_{\rm hot}}} 
\def\acool{{\alpha_{\rm cool}}} 
\def\Mhalo{{M_{\rm halo}}} 
\def\Mstar{{M_*}} 
\def\Emean{E_{\rm mean}} 
\def\Eesc{E_{\rm esc}} 
 
\def\lexpel{{\lambda_{\rm expel}}} 
\def\lexp200{{\lambda_{\rm expel, 200}}} 
\def\aexp{\alpha_{\rm exp}} 
\def\b200{{\beta_{200}}} 
\def\vw200{v_{\rm wind, 200}} 
\def\vwind{v_{\rm wind}} 
\def\vhalo{v_{\rm halo}} 
\def\vhot{v_{\rm hot}}
\def\vdisk{v_{\rm disk}} 
\def\ahot{\alpha_{\rm hot}} 
\def\esn{\epsilon_{\rm SN}}

\begin{document}

\title[What Shapes the Galaxy Mass Function?]
{What Shapes the Galaxy Mass Function? Exploring the Roles of Supernova-Driven Winds and AGN}

\author[Bower et al.] { 
\parbox[t]{\textwidth}{ \vspace{-1.0cm} R. G. Bower$^{1}$,
  A. J. Benson$^{2}$, Robert A. Crain$^{3,4}$ } \vspace*{6pt}\\
 $^{1}$Institute for Computational Cosmology, Department of Physics,
 University of Durham, South Road, Durham, DH1 3LE, UK.\\
 ({\it e-mail: r.g.bower@durham.ac.uk})\\
 $^{2}$Mail Code 350-17, California Institute of Technology, Pasadena,
 CA 91125, U.S.A.\\
 $^{3}$Centre for Astrophysics \& Supercomputing, Swinburne University
 of Technology, Hawthorn, Victoria 3122, Australia\\
 $^{4}$ Leiden Observatory, Leiden University, P. O. Box 9513, 2300 RA Leiden, the Netherlands\\
 \vspace*{-0.5cm}}

\maketitle

\begin{abstract} 
The observed stellar mass function (SMF) is very different to the halo mass function predicted by $\Lambda$CDM, and it is widely accepted that this is due to energy feedback from supernovae and black holes.  However, the strength and form of this feedback is not understood. In this paper, we use the phenomenological model GALFORM to explore how galaxy formation depends on the strength and halo mass dependence of feedback.  We focus on ``expulsion'' models in which the wind mass
loading, $\beta$ is proportional to $1/\vdisk^n$, with $n=0,1,2$ and contrast these models with the successful Bower et al.\  model (B8W7), for which $\beta
\propto 1/\vdisk^{3.2}$. A crucial development is that our code explicitly accounts for the recapture of expelled gas as the system's halo mass (and thus gravitational potential) increases. While models with high wind speed and mass loading result in a poor match to the observed SMF, a model with slower wind speed matches the flat portion of the SMF between $M_\star \sim 10^9-10^{11} \hMsol$. When combined with AGN feedback, the model provides a good description of the observed SMF above $10^9 \hMsol$. In order to explore the impact of different feedback schemes further, we  examine how the expulsion models compare with a further range of observational data, contrasting the results with the B8W7 model. In the
expulsion models, the brightest galaxies are assembled more recently,
and the specific star formation rates of galaxies decrease strongly
with decreasing stellar mass.  
{\rgb The expulsion models tend to have a cosmic
  star formation density that is dominated by lower mass galaxies at
  $z=1-3$, and dominated by high mass galaxies at low redshift. These
  trends are in conflict with observational data, but the comparison
  highlights some deficiencies of the B8W7 model also.  The
experiments in this paper give us important physical insight to the impact of the feedback process on the formation histories of galaxies, but the strong mass dependence of feedback adopted  in B8W7 still appears to provide the most promising description of the observed universe.} 
\end{abstract}

\section{Introduction}

A central issue in the study of galaxy formation is to understand the connection between the mass of galaxies and the mass of their associated dark matter haloes. This problem is not trivial: whilst the mass function of dark matter haloes predicted by the cold dark matter paradigm has a relatively steep slope ($d\log N/ d \log M \sim -0.9$ over the range of haloes mass relevant to the galaxy formation), the observed mass function of galaxies is characterised by a shallow Schechter function. The abundance of galaxies is nearly independent of stellar mass over the interval $M_\star \sim 10^8 -10^{10.5} \hMsol$, whilst at greater masses the abundance of galaxies declines exponentially. This raises two fundamental questions: i) why is the stellar mass function so flat below $\sim 10^{10.5}\hMsol$, when the abundance of haloes is such a strong function of mass, and ii) what physical processes are responsible for the exponential suppression of galaxies with mass greater than $\sim 10^{10.5}\hMsol$ \citep{white_frenk91,benson2003}?

In this paper we investigate the astrophysical processes that map the halo mass function to the galaxy stellar mass function (SMF). There are two common approaches to this problem. Arguably, the most appealing route is to numerically integrate the set of differential equations that describe the rudimentary astrophysical processes (e.g. gravity, hydrodynamics, radiative cooling, star and black hole formation) using a \emph{local} framework (e.g. a set of particles or a grid). This approach aims to evolve the cosmological density fluctuation power spectrum reflected by the cosmic microwave background using an {\it ab initio} description of the problem. This system of equations commonly over-predicts the universal abundance of stars --- a problem commonly known as the `over-cooling problem' \citep[e.g.][]{katz1996, balogh2001, keres2009, schaye2010}.

In order to get closer to a description of the observed universe, an additional set of processes (collectively known as ``feedback'') must be introduced to couple the energy, mass and metals returned by supernovae (SNe) and black holes (AGN) to the surrounding gas. {\rgb As well as
accounting for the stellar mass and gas content of galaxies, feedback
enriches the intergalactic medium with metals, potentially
resulting in an orthogonal set of observational constraints}. However, the wide range of scales involved in this problem (from AU scales to tens of Mpc) makes it infeasible to model these processes from first principles, forcing recourse to phenomenological, or `sub-grid', treatments (eg., \citet{springel_hernquist03, springel2005, okamoto2008, schaye_dv08, booth_schaye09}).

Because the properties of model galaxies are remarkably sensitive to the details of sub-grid models, an alternative approach is to establish a set of equations that describes the same astrophysical processes on \emph{macroscopic} 
scales, typically averaged over the physical scale of a galaxy. This approach is adopted by phenomenological, or `semi-analytic'\footnote{The term ``semi-analytic'' commonly used in the literature, but is
missleading. With modern computing power it is no longer critical that the resulting equations
can be solved analytically. The important distinction is that this type of model provides
a macroscopic description of the relevant processes. This approach is key to gaining 
insight into the problem. Such models are usually referred to as
``phenomenological'' in other areas of physics.} models, such as the GALFORM code considered here (\citet{cole2000, bower2006}, hereafter Bow06), see also \citet{kauffmann1993, hatton2003, delucia2006, somerville2008, guo2011}). For example, rather than computing the star formation rate at each point in a galaxy, such models typically compute the total rate of star formation over the entire galaxy, and assume that the rate is a simple function of global galaxy parameters
(such as the gas mass, disk size and rotation speed). This leads to an alternative set of differential equations, that provide a macroscopic description of the physics.  A phenomenological model then solves these relatively simple equations within the merging hierarchy of structure formation. Macroscopic treatments are by definition approximate, but their use is ubiquitous in all branches of physics. When applied carefully and within an understood range of validity, the resulting descriptions can lead to significant physical insight.

In this paper, we apply the GALFORM model to seek an understanding of the mass-dependence of the equations of feedback from galaxies. We focus initially on the role supernova-driven winds play in establishing the galaxy stellar mass function on scales below $\sim10^{10.5}\hMsol$. Although strong suppression of galaxy formation in low mass haloes is clearly necessary, our understanding of the process remains incomplete, largely because of the difficultly of calculating the effects of feedback from fundamental physical principles. The difficulty primarily arises because the interstellar medium in which SNe explode is inherently multiphase (and magnetized), and does not behave like an ideal gas. Thus, accurately calculating the impact of even a solitary supernova is extremely challenging, as the result depends strongly on the initial density of the medium into which energy is injected. As a sequence of SNe explode, a network of low density channels (or `chimneys') is established that allow the supernova ejecta to escape into the halo of the galaxy \citep{mckee_ostriker77, efstathiou2000, deavillez2007}.  However, without detailed calculations it impossible to estimate how the mass outflow rate and the specific energy of the outflowing material will depend on the star formation rate and the mass of the host galaxy.

In view of this uncertainty, all currently feasible simulations of galaxy formation parameterise the effect of supernovae and include it as a sub-grid calculation \citep{springel_hernquist03, dv_schaye08}.  In hydrodynamical simulations, winds are most commonly modeled by adding thermal energy to gas particles (or cells), or by giving gas particles a velocity kick.  Although adding thermal energy seems a promising route, the energy injected is easily radiated away if the temperature of the particles is too low \citep{katz1996}. Equally the injection of kinetic energy may drive thermodynamic shocks into the surrounding gas, and this thermal energy may also be radiated. These issues arise because the codes treat the complex multiphase interstellar medium (ISM) as a single fluid. Several schemes have been developed to circumvent the problem; one approach is to decouple the relevant particles from the numerical scheme for a period of time, either by preventing cooling for a period \citep{brooks2007}, or by decoupling the particles from hydrodynamical forces \citep{springel_hernquist03, okamoto_nemmen2008, oppenheimer2008}. Particles therefore retain the energy injected by feedback for a period of time, easing their escape from the dense interstellar medium of the galaxy. An alternative approach is to stochastically heat or kick the particles so that their energy is sufficiently high that their cooling time is long, or the shocks they drive are sufficiently strong that radiative losses are small \citep{dv_schaye08, creasey2011}. 

Because of the difficulty in interpreting the multi-phase nature of the outflow, the mass loading and velocity of winds are not yet strongly constrained by observation (but see \citet{martin2005, weiner2009, chan2010, rubin2011} for recent progress). Most hydrodynamical calculations have therefore adopted the simplest possible model, in which the mass loading and velocity of winds are independent of the system in which the feedback event is triggered. This also simplifies implementation of the feedback, since there is no requirement to estimate the environment of the ISM on the fly, for example by determining the mass of the dark matter halo, or the local gravitational potential. An exception is the momentum scaling model of \citet{oppenheimer2006}, where the wind parameters are set following a radiatively driven wind model \citep{murray2005}, and a number of similar schemes implemented in the OverWhelmingly Large Simulations \citep[OWLS][]{schaye2010}.

By contrast, most phenomenological (``semi-analytic'') models of galaxy formation assume that the parameters describing feedback adjust to ensure that the specific energy of outflows are matched to the binding energy of the halo. Conservation of total energy thus ensures that the mass loading of outflows is greater in dwarf galaxies than in larger systems \citep{dekel_silk86}. Such models are partially motivated by arguments pertaining to the porosity of the interstellar medium: analytic models \citep[e.g.][]{efstathiou2000} consider the formation of channels in the multiphase ISM and suggest that the porosity of the ISM is self-regulating and determined by the gravitational potential of the disc. In general, phenomenological models further assume that material expelled from the disc is recaptured on a timescale proportional to the dynamical time. This coefficient is allowed to be significantly larger than unity in some models in order to approximate the effect of expulsion of gas from the halo.

Current phenomenological models present a coherent picture for the formation of galaxies in a
cosmological context, and provide an excellent explanation of many diverse data-sets. Although the GALFORM model has been developed to explain the observed properties of galaxies (eg., Bow06, \citet{font2008, lagos2010}), it has also been shown to explain the X-ray scaling relations of groups and clusters \citep[][hereafter Bow08]{bower2008} and the optical and X-ray emission from AGN \citep{fanidakis2011}. The models can be used to generate convincing mock catalogues of the observable Universe \citep{cai2009} and applied to test the procedures used to derive physical parameters from astronomical observations, and to identify the priorities for the next generation of astronomical instruments. These successes have been driven by the inclusion of two key components of the models: i) galaxy winds that scale strongly with halo mass and ii) a ``hot-halo'' mode\footnote{This mode is often referred to as the `radio' mode, we prefer the term `hot-halo' as it emphasises that this type of feedback is assumed to only be effective when the cooling time of the halo is sufficiently long compared to the dynamical time. See \citet{fanidakis2011} for further discussion.} of AGN feedback in which the cooling of gas in quasi-hydrostatic haloes is suppressed. By altering the parameterisation of the feedback schemes here, we investigate whether the Bow06 choice is optimal, or whether alternative schemes can similarly reproduce the properties of observed galaxies.

A major advantage of semi-analytic models is that they enable the effects of sub-grid parametrization to be explored quickly and easily \citep{bower2010}. We exploit this aspect in this study, to explore the effect of various descriptions of outflows of material from galaxies. This requires us to generalise the GALFORM model to include the possibility that gas is expelled from the potential of dark matter haloes. 
Previous efforts to `calibrate' the semi-analytic method against numerical simulations have included gravity, hydrodynamics and cooling, but not effective feedback \citep{benson2001, helly2003, delucia2010}.
Part of the reason for this is the very different treatments of the winds from galaxies. Our extensions of
the code allow us to bring the two approaches into closer alignment, and we include a brief comparison with 
the Galaxies-Intergalactic Medium Interaction Calculation \citep[GIMIC][]{crain2009}. This provides
a series of relatively high resolution hydrodynamic simulations, featuring relatively high spatial and mass resolution ($\epsilon = 0.5\hkpc$ and $m_{\rm gas} = 1.4\times 10^6 \hMsol$ for the highest resolution realisations), and that trace a representative cosmological volume (four spherical volumes with comoving radius $18\hMpc$ and one with comoving radius $25\hMpc$). The simulations include radiative cooling \citep{wiersma_schaye2009}, and sub-grid prescriptions for star formation and the thermodynamics of the ISM \citep{schaye_dv08}, hydrodynamically-coupled supernova-driven winds \citep{dv_schaye08} and metal enrichment resulting from stellar evolution \citep{wiersma2009}. Some important successes include the X-ray scaling 
relations of $L_\star$ galaxies \citep{crain2010}, and the distribution of satellites and stellar halo properties of the Milky Way \citep{deason2011, font2011}. Our modified GALFORM scheme implements similar physics, and we show 
that its behviour is very similar to GIMIC. This opens a new avenue, allowing us to use GALFORM to better understand how the parametrisation of feedback impacts upon the formation and evolution of galaxies, and thus guide both the development of sub-grid treatments in hydrodynamical simulations, and the interpretation of observational data.

The structure of this paper is as follows. In \S2, we discuss the implementation of supernova-wind driven feedback schemes in semi-analytic models, and introduce a new scheme that allows gas and metals to be expelled from low mass haloes and later reaccreted when the binding energy of the halo has increased significantly. In contrast to many previous models, we do not assume that material is always reaccreted after a certain timescale, or that expelled material is always lost from the hierarchy. In \S3, we present a comparison of different feedback scalings, focusing on the difference between schemes that scale the parameters describing winds with halo mass, and those that adopt fixed parameters.  In \S4, we explore how feedback from supernovae can be combined with feedback from black holes in order to generate an exponential break in the mass function. In particular, we compare the effect of `hot halo' mode feedback with that of strong quasar-driven winds (similar to those considered by \citet{springel2005}). We consider a number of additional observational constraints in \S5. While several of the feedback schemes are able to reproduce the high mass part of the SMF, we show that the specific star formation rate and the downsizing of galaxy formation are important orthogonal constraints. We present a summary of our conclusions in \S6. Except where otherwise noted, we assume a WMAP7 cosmology, $\Omega_0=0.272, \Lambda_0=0.728, \Omega_b=0.0455, h_0=0.704, n_s=0.967, \sigma_8=0.81$ (Komatsu et al. 2011). Throughout, we convert observational quantities to the $h$ scaling of the theoretical model so that stellar mass are quoted in $\hMsol$ etc.

\section{Feedback in Phenomenological Models}

\subsection{Feedback as a Galactic Fountain}

We begin by reviewing the conventional approach to feedback in GALFORM, focussing on the implementation used in Bow06.  To summarise the key features, gas is expelled from the disk and assumed to circulate in the halo, falling back to the disk on roughly a dynamical time if the cooling time is sufficiently short. If the cooling time is long compared to the dynamical time, the halo is susceptible to a hot-halo mode of feedback if a sufficiently massive central AGN is present. The scheme results in a galactic fountain with material rising from the galaxy disk and later falling back. In the case of haloes with short cooling times, it may be more appropriate to picture the circulating material as cool clouds rather than as material heated to the halo virial temperature. 

We parameterise the rate at which gas is expelled from the disk into the halo as 
\begin{equation} 
\dot{M}_{\rm outflow} = \beta \dot{M}_* ,
\end{equation}
where $\dot{M}_*$ is the star formation rate and the macroscopic mass loading factor, and $\beta$ is 
\begin{equation} 
\beta = \left(\vdisk\over \vhot\right)^{-\ahot} .
\label{eq:beta} \end{equation}
Here, $\vdisk$ is the circular speed of the galaxy disk, the parameter $\vhot$ sets the overall normalisation of the wind loading, and the parameter $\ahot$ determines how the mass loading of the wind varies with the disk rotation speed. We will be careful to explicitly distinguish the macroscopic loading factor, $\beta$, which represents the loading of the wind as it escapes from the galaxy into the halo, from the sub-grid loading factor $\eta$, used in hydrodynamical simulations to represent the amount of interstellar medium (ISM) material heated
or kicked by the supernova remnant. If winds are hydrodynamically
coupled, the macroscopic mass loading is very likely to be significantly larger than $\eta$. Furthermore, the physical processes that determine $\eta$ are themselves highly uncertain and their treatment in numerical models is likely to be resolution dependent.

In the Bow06 implementation, material is modeled as leaving the disk with a specific energy comparable to the binding energy of the halo 
{\rgb (ie., $\vwind \sim \vhalo$)} . If we assume $\vdisk\sim\vhalo$ (where $\vhalo$ is the circular velocity of the halo at the virial radius), energy conservation then requires that $\ahot=2$ and that, in the standard implementation, no material leaves the halo completely. However, Bow06 found that this scaling did not sufficiently suppress the formation of small galaxies and a stronger scaling, $\ahot=3.2$, was adopted. This gives a good match to the observed $K$-band luminosity function. The stronger scaling implies that in small haloes supernovae couple more efficiently to the cold gas, resulting in a higher mass-loading of the wind.  Assuming a velocity $\sim\vhalo/\sqrt{2}$ is sufficient to drive the fountain, the fraction of the total supernova energy needed to power the fountain is 
\begin{eqnarray}
f_{\rm fountain} &&\sim {1\over 8} \left(\vhot\over430\kms\right)^{3.2}\\  
&& \times \left(\vhalo\over200\kms\right)^{-1.2} \left(\esn\over 2.5\times10^{49} \erghMsol\right)^{-1} \nonumber
\end{eqnarray} 
where $\esn$ is the energy produced by supernovae per unit mass of stars formed. Assuming $\esn = 2.5\times10^{49} \erghMsol$ \citep[appropriate for a Chabrier IMF,][]{dv_schaye08} sufficient energy is, in principle, available to power the fountain in haloes more massive than $\vhalo\sim65\kms$. Interestingly, \citet{font2011a} find that the properties of Milky Way satellite galaxies are best reproduced if the mass dependence of feedback 
saturates in such low mass haloes.

\subsection{Allowing for Mass Loss from the Halo}

In the revised implementation presented in this paper, we consider the case where the energy of the gas escaping the disk has systematically greater than the binding energy of the halo. {\rgb This treatment is necessary if we are to consistently account for
winds with outflow speeds that are independent of halo mass. From an observational
perspective, this type of wind may be required in order to account for the wide spread
distribution of metals in the universe.}
This type of wind was previously is considered in \citet{benson2003}, and we briefly review the implementation here. 

We introduce the parameter $\lexpel$ to reflect the excess energy of the wind relative to the binding energy of the halo. Specifically, we set the mean specific energy of the wind to 
\begin{equation} 
\Emean = {1\over 2} \lexpel \vhalo^2.  
\label{eq:emean_a2} 
\end{equation} 
Note that $\lexpel$ may be a function of halo mass (see section \ref{sec:general_model}). We parameterise the fraction of material that is able to escape using the cumulative energy distribution, $f_E(x)$, where $x = \Eesc / \Emean$ and $\Eesc$ is a measure of the wind specific energy needed to escape the halo. We will choose a monotonic function for $f_E$ so that $f_E(0)=1$, $f_E(1)=1/e\sim0.36$ and $f_E \rightarrow 0$ for large $x$. We will set the energy needed to escape the halo to $\Eesc = \vhalo^2$ (ie., we assume that the escape velocity is $\sqrt{2} \vhalo$). Clearly this is an over simplification, since the true escape velocity depends on the details of the potential, the launch radius of the wind, the terminal radius and the ram-pressure that the gas encounters.  While we adopt this scaling to give a simple interpretation the wind velocities, there is likely to be a systematic offset when comparing with hydrodynamical simulations. Since it is the ratio of $\Emean$ and $\Eesc$ that determines the result of a model, we could rescale the wind speeds quoted in this paper according to the new pre-factor. For convenience, we can represent $\lexpel$ as a wind speed, $\vwind$, where 
\begin{equation} 
\vwind = \lexpel^{1/2} \vhalo 
\label{eq:vwind} \end{equation} 
and we will refer to models by their wind speed; however, it should be
remembered that this is more accurately defined as the specific energy
of the wind, and we do not intend to imply that the wind necessarily
has a bulk outflow velocity of $\vwind$:  what really matters is
  the fraction of the mass of the outflow that escapes from the halo.
Combining equations \ref{eq:emean_a2} and \ref{eq:vwind}, a
significant fraction of the outflow will escape the halo if 
\begin{equation} 
\vwind > \sqrt{2} \vhalo. 
\end{equation} 
We will consider the halo mass dependence of $\vwind$ below, but it will be useful to normalise different models at a particular halo mass. For example, a fiducial halo with $\vhalo=200 \kms$, for which $\lexpel\equiv\lexp200=1$--4 corresponds to wind speeds, $\vwind = (200, 400, 600, 800) \kms$.

Material that is not expelled is added to the halo following the scheme described in Bow06.  This includes a delay proportional to the dynamical time before the material is allowed to cool again (see Bow06 for details). Material that escapes the halo may be later recaptured as the halo grows in mass. We implement this by scanning through descendant haloes in the dark matter merger tree and adding $f_E(\Eesc^{i}/\Emean) - f_E(\Eesc^{i-1}/\Emean)$ to the reheated gas mass at each step (where $\Eesc^{i}$ refers to the escape energy of the descendant halo at timestep $i$, and $i$ ranges from the step at which the energy is injected to the final output time, $i=0$).  Mass that is added to the halo becomes able to cool on the dynamical timescale (which we define as $G \Mhalo / v_{\rm halo}^3$).  It may not be able to cool if the cooling time is long and AGN feedback is sufficiently effective. The step in $\Eesc$ may be small if the halo grows only a little by accretion, or may be large if the halo is accreted to become part of a much larger structure.

Note that this scheme differs significantly from the superwind implementation of \citet{baugh2005}, in which expelled material is not considered for recapture.  Since the overall baryon fractions of clusters of galaxies are inferred to be close to the cosmic abundance, recapture must be an important part of the feedback process. Finally, we note that some semi-analytic models adopt a feedback scheme in which expelled material becomes available for cooling or star formation on a timescale that is much longer than the dynamical time.  This is an approximation to the superwind scheme that we have described here, but it is not accurate since it does not take the growth rate of the halo into account.

In order to fix on a scheme we must choose an appropriate form for the
(cumulative) distribution functions $f_E$. \citet{benson2003} chose an
exponential form, $f_E = e^{-x}$.  This leads to a broad spread of
wind particle energies. {\rgb On the basis of their observational data,
\cite{steidel2011} suggest that wind outflow is more sharply peaked},
and we also find that a more sharply peaked distribution better match the
results of hydrodynamical simulations. In the following models we will
assume $f_E = \exp(-x^6)$ in what follows.  The precise choice of
power is not important, however, and we obtain similar results for $f_E = \exp(-x^2)$.

\subsection{A Generalized Feedback Model} \label{sec:general_model}

In contrast with GALFORM, and largely because of a lack of observational motivation for any particular scaling, hydrodynamical simulations have mostly adopted the simplest case of assuming that the mass loading and velocity of winds are independent of halo properties. We can easily adapt the revised feedback implementation described above to investigate such a scheme in GALFORM, by explicitly including a halo-mass dependence in Eq.~\ref{eq:emean_a2}. 
We will adopt $\vhalo=200 \kms$ as a fiducial halo mass at which to compare the wind mass loading and wind speed for models with different $\aexp$ so that we express the mass dependence of the wind speed as $\lexpel = \lexp200 \left(\vhalo / 200 \kms\right)^{\aexp}$, where $\aexp$ is a dimensionless parameter that differentiates different feedback models. Combining this with Eq.~\ref{eq:emean_a2} and Eq.~\ref{eq:vwind} gives 
%\begin{equation} 
%\Emean = {1\over 2} \lexp200 \vhalo^{2+\aexp} (200 \kms)^{-\aexp}. 
%\label{eq:emean} \end{equation} 
\begin{equation}
\vwind^2 = \vw200^2 \left(\vhalo \over 200 \kms\right)^{\aexp+2}.
\end{equation}
For the mass loading we have
\begin{equation}
\beta = \b200 \left(\vw200 \over 200 \kms\right)^{-\ahot}
\end{equation}
In the case $\aexp=0$ we recover the wind specific energy scaling with the specific binding energy
of the halo. In the case $\aexp = -2$, the wind speed is independent of halo mass.  Unless otherwise 
stated, we will set 
\begin{equation} 
\ahot = \aexp + 2 
\end{equation} 
so that a fixed fraction of the supernova energy is used to drive winds in haloes of all masses (assuming 
$\vdisk\sim \vhalo$ in Eq.~\ref{eq:beta}).

Given a wind speed and mass loading normalisation, $\vw200$ and $\b200$, the code parameters are
\begin{equation} 
\lexp200 = \left(\vw200 \over 200 \kms\right)^2,
\end{equation} and
\begin{equation} 
\vhot = \b200^{1/\ahot} 200 \kms.
\end{equation} 
This allows simple comparison to older models. Note that the original GALFORM parameter, $\vhot$, expresses the mass loading, and is not a measure of the specific energy of the wind. To avoid this confusion in this paper,
we will use the macroscopic mass loading parameter $\b200$ to label models in what follows.
{\rgb The maximum available supernova energy sets a limit 
$\b200 (\vw200/200 \kms)^2 < 32$}. 

If $\aexp=-2$ ($\ahot=0$), the macroscopic mass loading, $\beta$, is
independent of the halo potential. This would be the case if the wind
were completely decoupled until it that had escaped from the halo. This would make it impossible to frame the feedback in terms of the standard GALFORM parameters, and for this paper, we will assume that there is always a small coupling between the wind loading and the halo mass.  We adopt $\aexp=-1.9$ ($\ahot=0.1$) as our minimum value.

While $\aexp=0$ and $-2$ are natural choices, there is no a-priori reason to adopt a particular value of $\aexp$, and we will consider $\aexp=-1$ ($\ahot=1$) as an intermediate value. For these parameters, the  
speed of the wind scales with $\sqrt{\vhalo}$, and its mass loading scales as $1/\vhalo$. 
In this case, the material expelled from smaller galaxies is more likely to escape the halo, but this is a weaker function of mass than in the superwind case discussed above. The scaling of the 
wind mass loading is similar to the momentum driven model used by \citet{oppenheimer2008},
but note that we will assume that the ratio of the total energy of the wind (not its total momentum) to the mass of stars formed is independent of halo mass.

\subsection{Parameter values}

\begin{table*}
\begin{tabular}{|l|c|r|r|r|r|r} 
\hline 
Model & & $\ahot$& $\b200$& $\vw200/\kms$& $\aexp$& $\acool$ \\ 
Bow06 & Bow06& 3.2& 17& --& --& 0.58\\
  \quad + W7 cosmology& B8W7& 3.2& 12& --& 0.0& 0.52\\
Superwind & pGIMIC& 0.1& 8& 275& -1.9& --\\
          & SW& 0.1& 8& 180& -1.9& (0.35)\\
Momentum Scaling & MS& 1.0& 8& 200& -1.0& (0.45)\\
Energy Scaling & ES& 2.0& --& --& 0.0& --\\
 \hline 
\end{tabular}
\caption{The table shows the models considered in this paper, and compares to the values adopted in Bow06.  Up dating the comsological paramaters
requires that we make small adjustments to restore a good match to the 
observed mass function.  We will refer to this model as B8W7. The following
models correspond to the optimal paramaters for the different feedback schemes considered in the text. The values in brackets indicate the values
adopted for the AGN ``hot halo'' of feedback in \S\ref{sec:radio_mode}. } 
\label{tab:runparam} 
\end{table*}

The parameters of the best fitting models are given in Table~\ref{tab:runparam}.  To simplify comparison with previous work, we have translated the feedback parameterisation in Bow06 ($\vhot=485\kms$, $\ahot=3.2$, $\lexp200=0$ and $\acool=0.58$) into to the more generalized parameters considered in this paper. Where parameter values are not explicitly given, we adopt those in Bow06 with the exceptions given below. First, we now adopt a background cosmology that is consistent with the WMAP 7-year results \citep{komatsu2011}. Secondly, we use a stellar yield of $p_{\rm yield} = 0.04$ in order to improve the match of galaxy colours as discussed in \citet{font2008}, and a default halo gas distribution with core radius of $0.025 r_{\rm vir}$ as discussed in Bow08. With these revisions we make small shifts in the standard feedback parameters in order to restore a good fit to the local $K$-band luminosity function. 

The parameters of the base-line model are given in the second row of Table~\ref{tab:runparam}.
We use $\vhot=430\kms$ and $\acool=0.52$, where $\acool$ determines the ratio of free-fall and cooling times at which haloes are taken to be hydrostatic (as opposed to being classified as `rapidly cooling') so that only when $t_{\rm cool}(r_{\rm cool}) > \alpha_{\rm cool}^{-1} t_{\rm ff}(r_{\rm cool})$ is the AGN feedback effective.
%\footnote{Equation (2) of Bow06 is incorrect; $\alpha_{\rm cool}$ should 
%be replaced by $\alpha_{\rm %cool}^{-1}$.}
These differences have little impact on the properties of sub-$L_*$ galaxies. Since we are initially concerned with the faint end of the luminosity function, we begin by disabling the AGN feedback scheme by setting $\acool=0$. This allows us to make a simple comparison to hydrodynamical calculations that do not include AGN feedback. 

We consider three supernova-driven feedback models following the feedback scheme discussed in the previous section. In what follows, we refer to these as `superwind' (SW), `momentum scaling' (MS) and `energy scaling' (ES) models. This nomenclature reflects how the mass loading scales with halo mass, corresponding to $\ahot = 0.1, 1$ and~2.  For each case, we typically consider six values of the mass loading at $\vhalo=200 \kms$, $\b200 = 1, 2, 4, 8, 16$ and~32, and five values of the wind launch velocity at this halo mass, $\vw200 = 100, 200,400,600$ and~$800\kms$.  For the superwind and momentum scaling models, optimal values of the feedback parameters have been chosen to provide a reasonable match to the SMF above $\Mstar \sim 10^9 \hMsol$ (a particularly good fit was not possible for the ES scaling) and the corresponding parameters are given in Table~\ref{tab:runparam}. We also consider a model intended to replicate the supernova-driven winds implemented in the GIMIC hydrodynamical simulations \citep{crain2009}. These adopt a fixed sub-grid wind mass loading of $\eta=4$, and a launch wind velocity of $600\kms$. This requires $\sim 80$ percent of the available supernova energy being used to drive the wind. We find that we can best reproduce the resulting stellar mass function produced by the simulation if we adopt a somewhat lower macroscopic wind speed, $\vw200=275 \kms$, and higher mass loading, $\b200=8$ (although a broad range of parameters give similar results). Since the winds in the simulation are always hydrodynamically coupled, the increased mass loading of the wind is not surprising. We will refer to this `pseudo-GIMIC' model as pGIMIC below.

While we initially consider the models in the absence of AGN feedback, we will later adjust the AGN feedback parameter $\acool$ in order to set the mass scale of the break in the luminosity function. For this we use the AGN feedback scheme of \cite{bower2008}. This allows material to be expelled from the central regions of haloes through AGN heating, and results in the baryon fraction in galaxy groups being much lower than the cosmic average. This provides a much improved match to the observed X-ray properties of these systems. The values of $\acool$ that give a good match to the observed luminosity function are given in brackets in the table.  The ES model could not be adjusted to give a sufficiently flat faint end slope to the SMF, and we do not consider the role of AGN in this model.

\section{Galactic Wind Models}

\subsection{Conventional GALFORM Winds}

\begin{figure} 
\begin{center}
\includegraphics[scale=0.4]{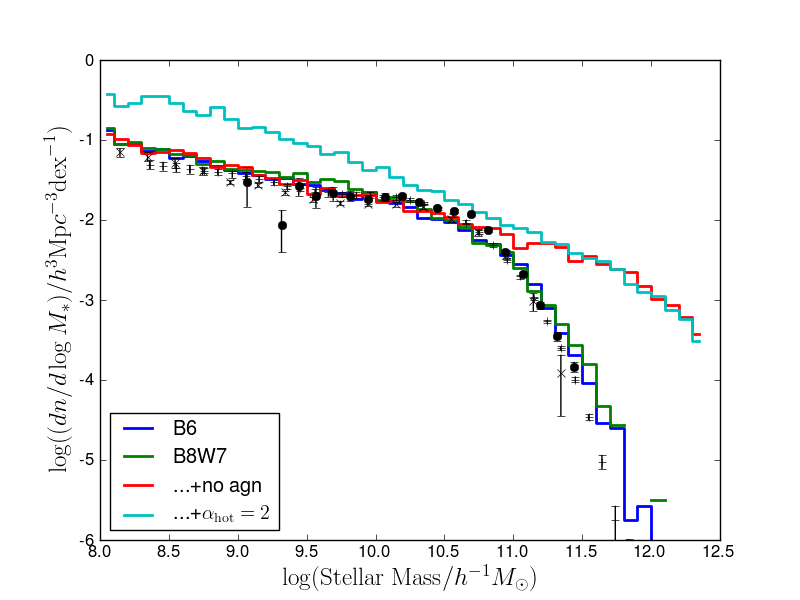} 
\end{center} 
\caption{Comparison of the stellar mass function of the Bow06 model (blue line) with the base-line B8W7 model used in this paper (green line). This is based on the WMAP7 cosmology and includes AGN ``hot halo'' feedback following Bower et al.\ 2008.  The two models are almost indistinguishable. To illustrate the importance of AGN feedback, we show the effect of turning off the AGN feedback (red line). We also show the effect of then adopting the $\ahot=-2$ (cyan line). For comparison, observational data are shown as black points with error bars. The data is taken from \citet{bell2003} (circles) and \citet{li_white09} (crosses).} \label{fig:Bow06} 
\end{figure}

We use the galaxy SMF as the starting point for comparing the models and the data. We use the determination of the stellar mass function by \citet{bell2003} and \citet{li_white09}, correcting 
the IMF to the Kennicut parameterisation. We convert the observational data to the $h^{-1}$ dependency on the Hubble parameter of the the theoretical models. The data  are compared with the Bow06 model, and the baseline 
B8W7 model in Fig.~\ref{fig:Bow06}. 
Both models assume that supernova-driven winds do not escape the parent 
halo; {\rgb instead the material ejected from the disk circulates in the halo and returns on the dynamical timescale if the cooling time is short}.
The two models are almost indistinguishable, since changes in the background cosmology and the AGN feedback implementation have been intentionally compensated for by small changes in the feedback parameters.
As expected, both provide a good description of the observational data.

When we consider alternative wind descriptions, we will not initially consider AGN feedback. To establish a baseline for the comparison in the absence of AGN feedback, we also show the B8W7 model with AGN feedback disabled (by setting $\acool=0$). This is shown as a red line in the figure. The roll-over of the galaxy luminosity function is almost non-existent in this model, while the SMF is largely unaffected below $10^{11}M_{\odot}$. This is encouraging, since it shows that the two processes involved in matching the shape of the galaxy luminosity function (ie., eliminating the overabundance of galaxies of fainter than $\sim 10^{10}\hMsol$ and reducing the abundance of galaxies above the break in the mass function) can be separated. We therefore focus our initial discussion on the modes of supernova-driven feedback, and consider models that do not include an AGN feedback component.

The Bow06 and B8W7 models achieve a good fit to the abundance of low mass galaxies because of the very strong mass dependence of the feedback ($\ahot=3.2$). For comparison, we show a model with feedback parameters that might be considered to be better motivated by theory. By selecting $\ahot=2$, the wind speed is tuned to the binding energy of the halo. As can be seen, this results in a somewhat steeper faint end-slope and a relatively poor match to the observed stellar mass function. In the following section we consider in more detail how the choice of feedback scheme affects the mass function, and how the wind parameters can be optimised to improve the fit.

\begin{figure*} 
\begin{center} 
\includegraphics[scale=0.4]{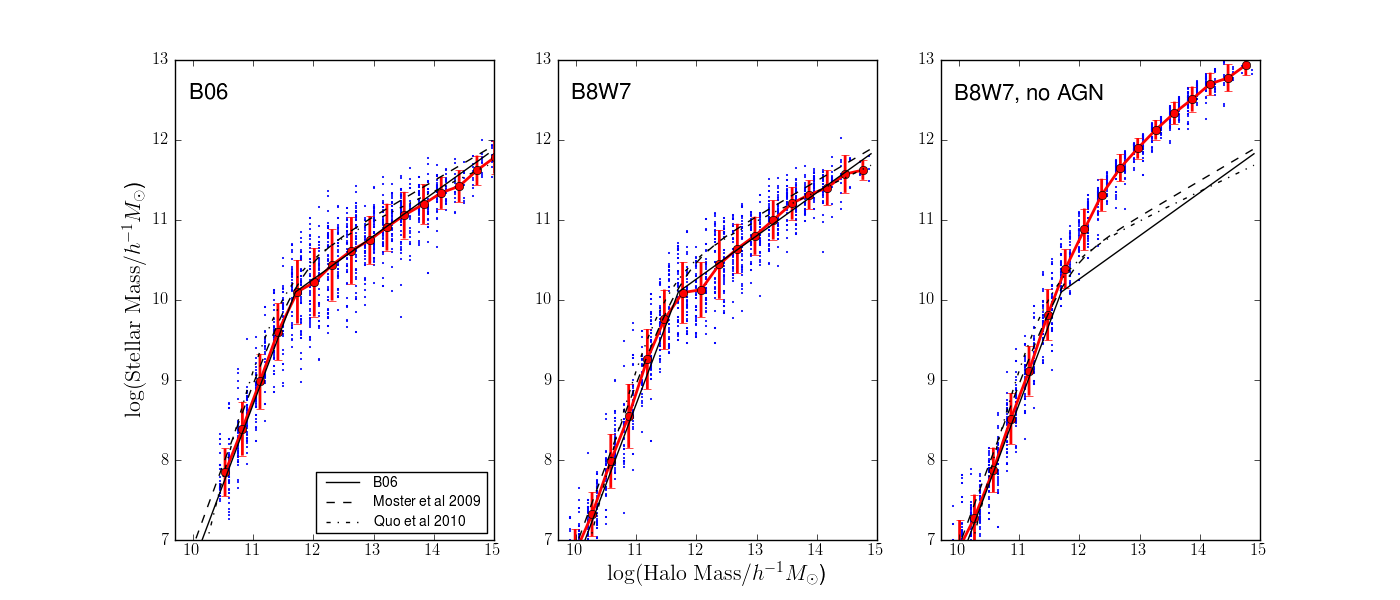} \end{center} \caption{The $M_*$ -- $\Mhalo$ relation for central galaxies. In this figure the panels show (from left to right) Bow06; the base-line model B8W7; and B8W7 without AGN feedback. Blue points show a random selection of haloes from the model, while the solid red line shows the median, and error bars show the 
$1\sigma$ scatter. The solid black line shows an empirical fit to the Bow06 model, while black dashed and dot-dashed lines show halo abundance matching models from \citet{moster2010} and \citet{guo2010} respectively.  The models in the first two panels include AGN feedback and successfully match the observed stellar mass function. These are well characterised by a broken power-law. Although the first two panels have slightly different background cosmology, the relation between halo mass and stellar mass is similar. Disabling AGN feedback, as shown in the last panel, results in a much weaker roll-over of the relationship. } \label{fig:mstar_mhalo_bow06}
\end{figure*}

Although the SMF provides a good way to compare the results of different feedback schemes with observation, it is far from simple to interpret the changes in terms of the effect of the different feedback schemes.  For example, increasing the effectiveness of the feedback scheme shifts galaxies to lower stellar mass, and so only affects the normalisation of the mass function indirectly. The suppression of the normalisation arises both because of the lower abundance of the haloes of greater mass and because of the range of halo masses that contribute galaxies of a particular stellar mass. A better way to understand the effect of the schemes is therefore to plot the stellar mass of the central galaxy as a function of halo mass. Since the scatter is not strongly constrained observationally \citep{moster2010}, we use the relationship found in the Bow06 and B8W7 models as a best guide to the relationship expected in the real Universe. We can then understand how various feedback schemes affect this relationship, and compare with constraints yielded by abundance matching observations with theoretical subhalo mass functions. {\rgb3 Of course, the parameters chosen in Bow06 and B8W7
are not unique, and other parameter combinations can give similar
quality fits to the mass function and other datasets, as we have shown in 
\cite{bower2010}, but the models to provide a well documented
starting point for our comparison of different feedback schemes.}

The first panel of Fig.~\ref{fig:mstar_mhalo_bow06} illustrates the dependency of central galaxy stellar mass on halo mass for the Bow06 model. The red line shows the mean relation. The dashed black line shows a broken power law approximation to the model, described by 
\begin{eqnarray}
\log \Mstar &=& 10.1 + 0.54 (\log(\Mhalo) - 11.7) \nonumber\\  
&& \qquad\qquad\qquad {\rm if} \log(\Mhalo) \ge 11.7 \nonumber\\  
&=& 10.1 + 2 (\log(\Mhalo) - 11.7) \nonumber\\  
&& \qquad\qquad\qquad {\rm if} \log(\Mhalo) < 11.7.
\end{eqnarray} 
At the break in the curve, 14 percent of the baryons in the halo have
been converted into stars. Note that there is considerable scatter
about these relations in the model. We repeat this relation in all
plots so that they can be compared easily. {\rgb The error bars indicate the
$\pm1\sigma$ range of the model galaxies.} Close to the break, the
scatter in the model exceeds an order of magnitude. 
We supplement the empirical approximation with relations from
\citet{moster2010} and \citet{guo2010}. {\rgb The relations shown are derived from
matching the abundance of sub-haloes in $N$-body simulations to
observational data, assuming that the scatter in the relation is
negligible. The models are based on $(\Omega_m,
\Omega_\Lambda, h, \sigma_8 , n_s)	= (0.26,	0.74,	0.72,
0.77,	0.95)$ and $(0.25, 0.75, 0.73, 0.9, 1)$ cosmologies
respectively, but note that the differences in the predicted abundance
of $10^{10}-10^{12}\hMsol$ haloes are small. Thus the relations are
similar for stellar masses below $10^{10}\hMsol$ but are offset from
the Bow06 relation at high masses due to the large scatter
about the mean relation. Because of the steep break in the mass
function, scatter boosts the abundance of massive galaxies relative
to a relation without scatter (see \citet{moster2010} for further
discussion), and thus the scatter and the normalisation of the 
high mass $\Mstar - \Mhalo$ are tightly correlated.}

The second panel of Fig.~\ref{fig:mstar_mhalo_bow06} shows the
baseline B8W7 model. The scatter in $\Mstar$ at a given halo mass is
reduced compared to Bow06, although it is still larger ($\sigma=0.3$ dex) than the scatter explored by \citet{moster2010} (up to 0.15 dex), particularly around the break in the relation. 
The final panel shows the effect of {\it disabling} AGN feedback in the B8W7 model. The power-law relation now extends to higher mass before slowly rolling over as the result of the increasing cooling times of massive haloes. The scatter in the relation around $\Mhalo \sim 10^{12}\hMsol$ is now much reduced. This arises because the efficacy of AGN feedback in this model has a strong dependence on the accretion history of haloes (see Bow08 for further discussion).

{\rgb In summary, the B8W7 model provides a match to the observed stellar mass function
due to the very strong halo mass dependence of the wind mass loading and the suppression
of cooling in haloes with relatively long cooling times. This paper investigates
whether models with more general feedback schemes can achieve a similar success.}

\subsection{Superwind Models with Fixed Wind Speed and Mass Loading}

\subsubsection{Effect of wind parameters}
\label{sec:effect_of_wind_params}

We now consider superwind (SW) models, in which the mass loading and velocity of winds are (almost) independent of the halo mass.  This mimics the schemes that are commonly adopted in hydrodynamical simulations. We begin by contrasting the results with the baseline B8W7 model. 

\begin{figure} 
\begin{center} 
\includegraphics[scale=0.4]{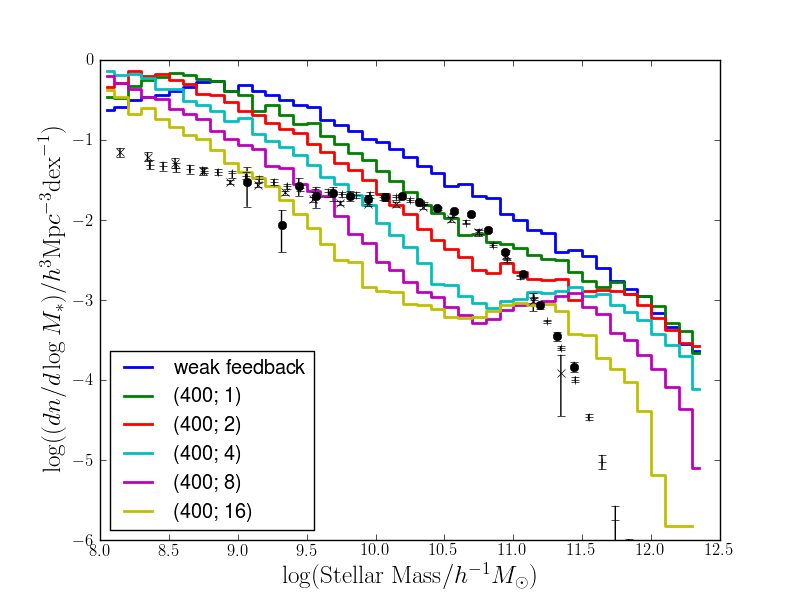} \end{center} \caption{Effect of mass loading in the superwind (SW) model. These models (solid lines) have a wind velocity of $\vw200=400\kms$, and mass loading $\b200 = 1,2,4,8,16$. We also include a model with low mass loading ($\b200=1$) and low speed ($\vw200=100 \kms$) to show the effect of suppressing the feedback channel (blue line). 
With low mass loading, the faint end of the function is extremely steep. Increasing the mass loading suppresses the normalisation of the low mass part of the mass function, but tends to introduce a 
noticeable dip in the SMF if the loading is high. With a high mass loading
also suppresses the abundance of the highest mass galaxies, but shape
of the mass function does not match the observational data. Since AGN feedback has not been included, the cut-off seen here is the result of the 
long cooling times in larger haloes. 
} \label{fig:wind_mloading} 
\end{figure}

Fig.~\ref{fig:wind_mloading} shows the effect of varying the mass loading for a fiducial wind speed of $\vw200=400 \kms$.  For comparison, the blue curve shows the effect of low mass loading and low wind speed such that supernova driven feedback is ineffective.
Increasing the mass loading reduces the normalisation of the mass function below $M_*$, but the power-law dependence at low galaxy luminosities becomes steep, increasing the discrepancy with observations.  As the mass loading increases above $\b200=4$, feedback takes a `bite' out of the mass function. This can be understood as a transition in the effectiveness of feedback. In high mass haloes ($\vhalo>\vw200$) material falls back onto the central galaxy on the dynamical timescale, while in low mass haloes it is expelled and ceases to be available to fuel star formation. The timescale for the return of expelled material therefore makes a transition when the two speeds are equal \citep[see][]{oppenheimer2008}. (The effect is most clearly seen by plotting the stellar mass of galaxies against their halo mass, as we discuss below.) Although increasing the mass loading tends to suppress the abundance of $M_*$ galaxies, the faint end slope of the mass function is always much steeper than the observations. It is not possible to improve the fit to the mass function by adjusting this parameter.

Although the abundance of bright galaxies greatly exceeds the observations, a roll-over is evident when the high mass loading. As was the case where AGN were disabled in the B8W7 model, the roll over is driven by the long cooling times of high mass haloes. In this situation, the bottle-neck is the cooling time of the material in the halo, and the star formation rate is
proportional to the inverse of the wind mass loading.

\begin{figure} 
\begin{center} 
\includegraphics[scale=0.4]{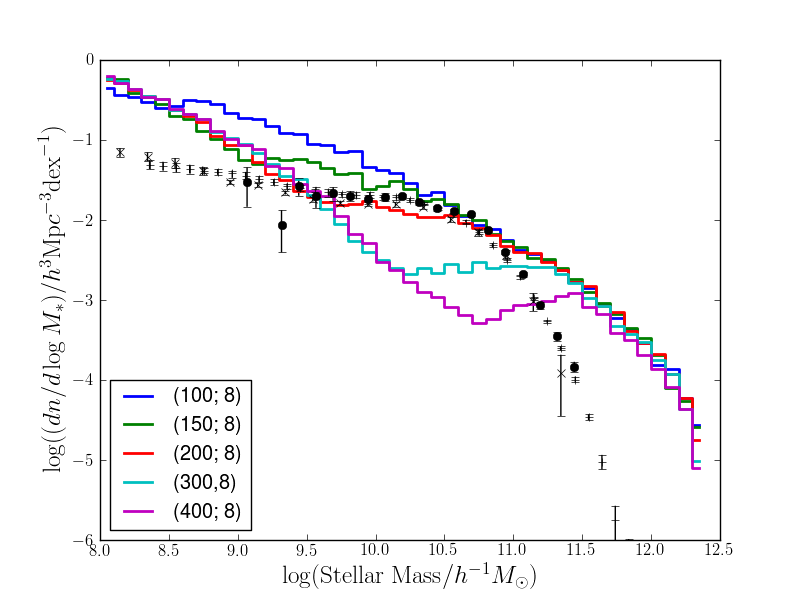} \end{center} 
\caption{The effect of changing the wind speed in the superwind (SW) model. These models have a fixed mass loading $\b200 = 8$ and varying wind speed $\vwind=100-400 \kms$. The figure shows how changing the speed of the wind adjusts the location and depth of the dip in the SMF, although the faintest and most massive galaxies are almost unaffected. An intermediate wind speed has the effect of producing a flat portion of the luminosity function.} 
\label{fig:wind_ewind} 
\end{figure}

The effect of changing the wind speed is explored in
Fig.~\ref{fig:wind_ewind}, where we show the mass function obtained
by varying $\vw200=100 - 400 \kms$ at a fixed mass loading
of $\b200=8$.  At a wind speeds greater than $200\kms$, the mass
function begins to resemble the observational data with a flat slope
around the knee of the luminosity function. This is encouraging: we
infer that that a suitable choice of wind parameters enables a match
to the properties of $10^{9.5}$ - $10^{11} \hMsol$ galaxies to be
obtained. However, while the model matches the observational data down
to $\Mstar \sim 10^{9.5}$, the number of galaxies rises rapidly at
lower masses, and an additional feedback mechanism would need to be
introduced to explain the low abundance of dwarf galaxies.  Since the
halo masses of these galaxies are sufficiently high that they are
unlikely to be affected by photo-heating
\citep{crain2007,okamoto2008}, the
only option would be to explore winds that scale with halo mass. 
{\rgb
On the other hand a high abundance of faint galaxies (at $z>6$) would provide  an abundant source of ionising photons to drive the re-ionisation of the
universe \citep{benson2006,jaacks2011}. If the mass-dependent scheme 
suppress the formation of pre-reionisation small galaxies too dramatically
there will not be sufficient photons to re-ionise the universe. The constraint is quite weak, however. Even with the strong halo mass dependence winds
in the B06 model, we find that it is sufficient to assume that feedback saturates when $\vhalo<65\kms$  in order to provide the
necessary ionising flux \citep{font2011a}.}

{\rgb  In summary, with suitable choice of parameters, the SW scheme offers an attractive explanation for the flat portion of the SMF in the range $10^{9.5}$ - $10^{11} \hMsol$. The model, however, predicts that lower mass galaxies will be more abundant than observed. In contrast, the strong B8W7
halo mass dependence of feedback in the B8W7 model results in a flat stellar
mass function down to below $10^8\hMsol$.}

\subsubsection{Comparison with Hydrodynamical Simulations} \label{sec:gimic_comparison}

\begin{figure} 
\begin{center} 
\includegraphics[scale=0.4,angle=0]{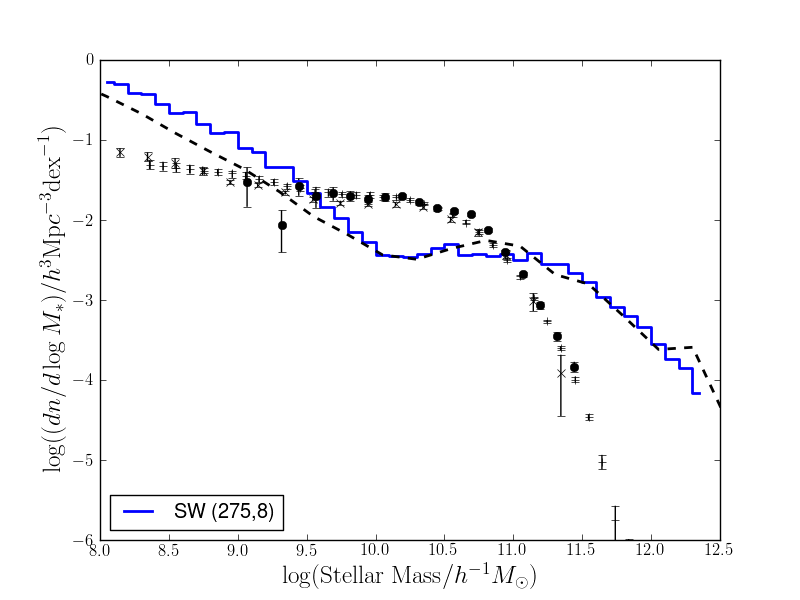} 
\end{center} 
\caption{Comparison with the SMF of the GIMIC hydrodynamic simulation (dashed line) and the GALFORM model (pGIMIC) with $(\vw200,\b200) = (275,8)$ (solid blue histogram).  The GALFORM parameters have been chosen to 
reproduce the position and amplitude of the dip seen in the hydrodynamic simulation. The GIMIC wind parameters are $(600, 4)$, but hydrodynamical coupling is expected to result in additional macroscopic mass loading and lower effective wind speeds. Observational data is reproduced from Fig.~\ref{fig:Bow06}. } 
\label{fig:gimic_starmf} 
\end{figure}

{\rgb
The SW feedback scheme is similar to the approach adopted in many
hydrodynamic simulations, and it is interesting to briefly compare the results in order to gain confidence that the phenomenological description we use
appropriately represents the physics of a full hydrodynamical
treatment. We first compare with the Galaxies and the Intergalactic Medium Calculation (GIMIC, \citet{crain2009}).
The highest resolution realisations of these simulations have gas particles with mass $1.45\times10^6 \hMsol$, and a softening length of $0.5\hkpc$. 
This is sufficient to resolve the onset of the Jeans' instability
in galactic disks while at the same time allowing reconstruction of a representative cosmological volume. Feedback is implemented by imparting kinetic energy to stochastically chosen neighboring particles of newly-formed stars. The kicked particles remain hydrodynamically coupled at all times. The simulations adopt sub-grid wind parameters $\eta=4$ and $\vw200=600 \kms$. This choice was motivated by observations suggesting that the wind speed was independent of halo mass
(see \citet{martin2005} for discussion) and by requiring that the total stellar mass density matched the observed universe.
Alternative choices are explored in the OWLS simulations \citep{schaye2010}. These parameters determine the input properties of particles. Since the simulation is fully
hydrodynamic (so that wind particles remain hydrodynamically coupled to the surrounding
gas particles), we should not expected them to directly translate into the macroscopic wind parameters used in GALFORM. 
}

In Fig.~\ref{fig:gimic_starmf} we compare the SMF from GIMIC with
the SW model  (a full comparison of individual galaxy merger trees will be
presented in a future paper). It is important to note that the GIMIC simulations did not include AGN feedback, and were constrained to matching the observed stellar mass density, rather than the portion of the mass function below stellar masses of $10^{11}\hMsol$. 
In order to run the GALFORM model, we revert to the cosmological parameters used in Bow06 (both models are based on the Millennium simulations, \citet{springel2005}).  With suitable choice of wind normalisation ($\vw200=275\kms$ and $\b200=8$), the GALFORM code reproduces the hydrodynamic mass function well. Although the GALFORM wind has lower speed and somewhat higher mass loading, it should be remembered that these are the effective macroscopic wind parameters. As was shown by \citet{dv_schaye08}, the ram pressure induced by the hydrodynamic coupling of winds tends to slow the outflow and increase its mass loading as it leaves the disk of the galaxy.

\citet{oppenheimer2008} (hereafter Op08) also present similar models with which we can compare.
These simulations have significantly lower mass resolution than GIMIC, with gas particle masses of up to $1.5\times10^7 \hMsol$ and a softening length of $1.9\hkpc$. Consequently, these hydrodynamic simulations do not attempt to resolve physics within galaxies. Star formation is implemented following the sub-grid scheme of \citet{springel_hernquist03}: winds are implemented kinetically, but the affected particles are decoupled from hydrodynamic forces until the surrounding density is low. 

Op08 consider three distinct models. A high wind speed model, a model
with much lower wind speed, and another in which the wind speed (and
mass loading) scale (inversely) with the local halo velocity
dispersion. Their strong wind model ($\vw200,\b200 = 680,2$) produces
results that are similar to those of the GIMIC simulations. However,
the `slow wind' model ($\vw200,\b200 = 340,2$) provides a good match
to the observed mass function over the range plotted in their paper. There
are three regimes of the SMF for the slow wind. A
steep slope at low mass, flat around $10^{10}\hMsol$ and then steep
again at higher masses. As we have seen, GALFORM can reproduce this
behaviour if the wind velocity is low ($200\kms$) and the mass loading
somewhat higher (between 4 and 8). These have comparable total wind
energy to the Op08 models. In low mass haloes, even the high wind
loading considered does not sufficiently suppress star formation,
compared to the observations.  At intermediate
masses, the slope is roughly flat as the wind becomes less effective
and eventually stalls.  Then at very high mass, cooling becomes
inefficient and the slope steepens. Obviously, as with GIMIC, the
match to the observed SMF at such high masses is
poor because the simulations do not include AGN feedback (but see
\citet{gabor2011}). We will consider models in which the wind parameters scale with the properties of the halo in Section~\ref{sec:MS}.

{\rgb In summary, this brief comparison shows that the expulsion scheme implemented in
our phenomenological model describes the effects seen in
hydrodynamic simulations well. By using these models to better explore
the parameter space of galaxy feedback, we can create a closer connection between phenomenological models and fully hydrodynamic simulations.
Comparison with the observational data highlights two important issues: firstly, the steep slope
of the mass function below  $\Mstar = 10^{10}\hMsol$ and secondly the
over-abundance of galaxies more massive than  $\Mstar =
10^{11.5}\hMsol$. In the following sections, we will explore how these
discrepancies can be resolved by introducing more feedback
schemes that scale with halo mass, and including feedback from AGN.}

%\begin{figure*} 
%\begin{center} 
%\includegraphics[scale=0.8,angle=0]{figures/gimic_starmf_evo.png} 
%\end{center}
%\caption{The evolution of the SMF for the pGIMIC model %$(\vw200,\b200) = %(275,8)$. The $z=0$ mass function is reproduced in each %panel for comparison %(dashed line).  For comparison, we show observational %data for the high %redshift mass function from a variety of sources (see text %for details). In %the first 
%panel the local data is reproduced from Fig.~\ref{fig:gimic_starmf}).  As was %found in the hydrodynamical GIMIC simulation, the model provides a fair %description of the $z=1$--3 data, even though it fails to match 
%the mass function at lower redshift.} 
%\label{fig:gimic_starmf_evo} 
%\end{figure*}

\subsubsection{Stellar Mass as a Function of Halo Mass}

\begin{figure*} 
\begin{center}
\includegraphics[scale=0.4,angle=0]{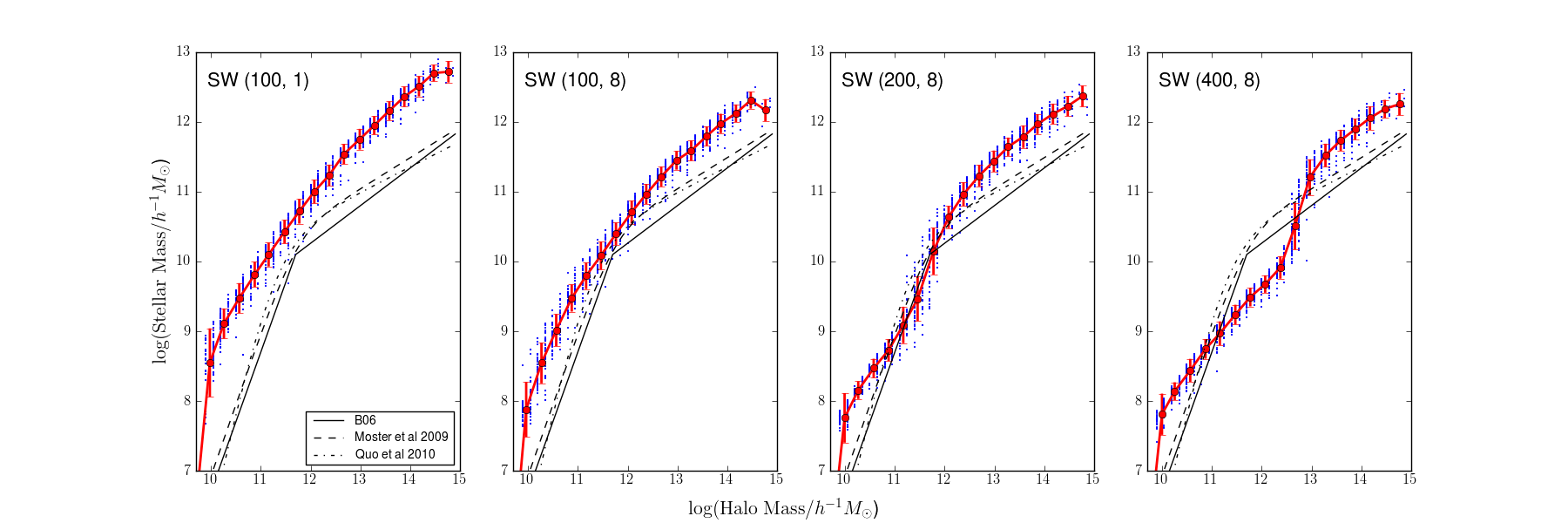}
\end{center}
\caption{Comparison of the $M_*$ -- $\Mhalo$ relation for SW models with various parameters. The first panel
shows a model with weak feedback $(\vw200,\b200) = (100,1)$. The remaining panels show 
$(\vw200,\b200) = (100,8), (200,8)$ and $(400,8)$ (from left to right) to illustrate 
the effect of increasing the wind speed at a fixed mass loading.
The SMFs corresponding to these models can be seen in 
Fig.~\ref{fig:wind_ewind}, the effects seen in the SMF are more readily interpreted 
in terms of the amplitude and slope of the $M_*$ -- $\Mhalo$ relation.
Black lines show the expected relationship derived from on observational data (see Fig.~\ref{fig:mstar_mhalo_bow06}).}  
\label{fig:mstar_mhalo_fixw}
\end{figure*}

In order to better understand how the feedback scheme can shape the SMF, it is useful to examine the relation between the halo mass and the stellar mass of central galaxies. We presented the relation for the Bow06 model in Fig.~\ref{fig:mstar_mhalo_bow06} and showed that the relationship can be characterised by a broken power law. Fig.~\ref{fig:mstar_mhalo_fixw} illustrates the effect of changing the feedback scheme to the SW model.  

The first panel shows the effect of including only minimal feedback,
$(\vw200,\b200) = (100,1)$. The discrepancies compared to the
empirical relationship (black lines) are evident. At all halo masses,
the associated stellar mass is too high, and the relation shows little
change of slope. This model corresponds to the weak feedback (blue)
line in Fig.~\ref{fig:wind_mloading}. At a given stellar mass, the
galaxies are formed in lower mass haloes than indicated by the
observed relation. These haloes are much more abundant, and thus the SMF is normalised too high. The slope of the relation is also in clear disagreement and this results in the overly steep faint end slope of the predicted mass function.

The remaining panels illustrate the effect of increasing the wind speed at a fixed mass loading. In the second panel, $(\vw200,\b200) = (100,8)$. This model is shown as the blue line in Fig~\ref{fig:wind_ewind}. The outflow has a low speed, so little mass escapes the halo, but the high mass loading results in effective suppression of galaxy stellar mass. As a result, the model matches the normalisation of the knee of the SMF well, and this is reflected by the $M_*-\Mhalo$ relation coming close to the kink of the observed relationship. However, several discrepancies from the observed relationship remain clear. In particular, the relation shows little change of slope. While the relationship at high stellar mass can be improved with AGN feedback, the relation at lower stellar masses is too shallow. As a result, galaxies of a given stellar mass are over abundant compared to the observed relation, as is evident in Fig~\ref{fig:wind_ewind}.

The third and fourth panels show the effect of increasing the mass loading in the model, $(\vw200,\b200) = (200,8)$ and  $(400, 8)$.
These are shown as the green and purple lines in
Fig.~\ref{fig:wind_ewind}. The increasing the wind speed creates a
kink in the $M_*$--$\Mhalo$ relation, with the stellar mass formed in
haloes around $10^{11}\hMsol$ being very strongly suppressed. In the kinked region, the steepening of the $M_*$--$\Mhalo$ relation means that a particular halo mass contributes to a wide spread of stellar masses, resulting in a suppression of the mass function normalisation. By suitable adjustment of
the parameters, the suppression can be tuned to create a flat portion
of the SMF. We will exploit this in
Section~\ref{sec:radio_mode}. {\rgb Below the kink, however, the slope
of the  $M_*$--$\Mhalo$ relation is much shallower than that seen in
the B8W7 model (Fig.~\ref{fig:Bow06}) and the slope of the mass
function is therefore inevitably steeper than the observational data.}

The kink is created by the wind stalling at a particular halo mass so that material no longer escapes the halo and falls back on the dynamical timescale. Increasing the speed of the wind shifts the region of the kink, but leaves the relations at high and low halo masses unchanged. This is reflected in the SMF, with the wind speed effecting a transition region between unchanging abundances of high and low mass galaxies. The transition between the regimes appears to become steeper for higher wind energies, resulting in a noticeable dip in the mass function.

\subsection{Momentum Scaling Models} \label{sec:MS}

\begin{figure*} 
\begin{center} 
\begin{tabular}{cc} 
\includegraphics[scale=0.4]{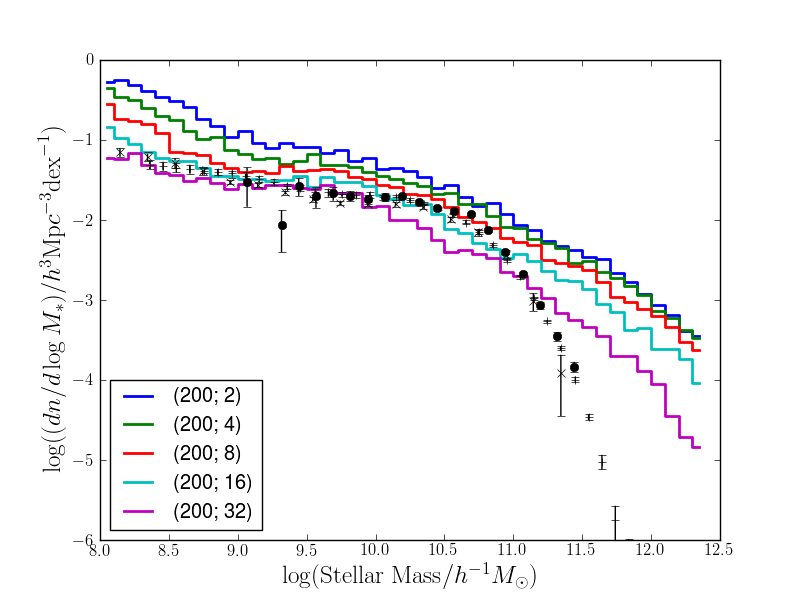} & 
\includegraphics[scale=0.4]{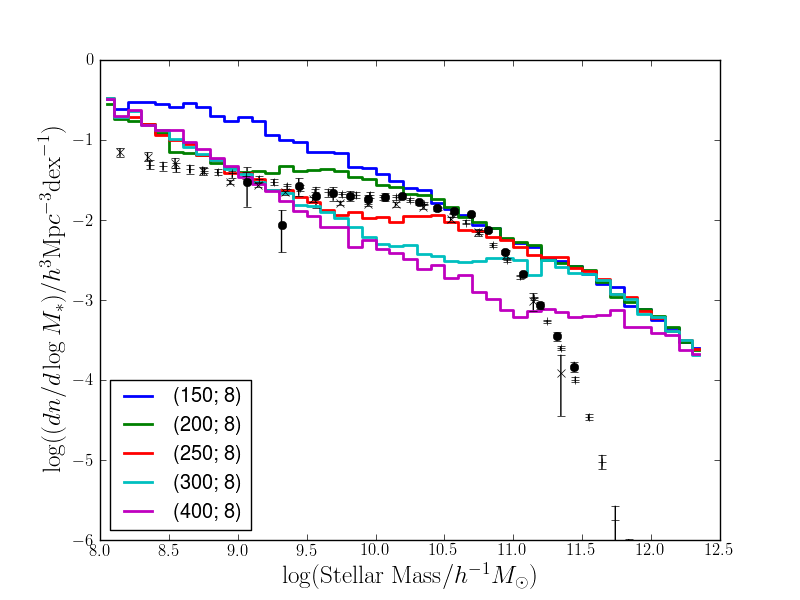} \\ 
\end{tabular} 
\end{center} 
\caption{The effect of using the ``momentum'' scaling (MS) model ($\ahot=1$).  The first panel shows a wind speed normalisation of $\vwind=200\kms$, with varying mass loading, $\b200=2-32$. For a low wind speed, a high normalisation of the mass loading effectively suppresses the formation of low mass galaxies leading to an encouraging match to the a large portion of mass function. The second panel shows a fixed wind loading of 8 and varying normalisation of the wind speed, $\vwind=100-400 \kms$. Increasing the wind speed has the undesirable effect of creating a dip in the mass function. } \label{fig:momdriven}
\end{figure*}

While Op08 find that the slow wind model fits the observed data over a
range from $10^{9.5}$ to $10^{11}\hMsol$, their preferred model is one
in which the feedback parameters vary with halo mass. Their preferred
scheme is intended to mimic the effect of a momentum driven wind
\citep{murray2005}, and the ratio of total momentum to mass of stars
formed is held fixed. We adopt a similar scheme, scaling the wind mass
loading inversely with the circular speed of the disk (ie.,
$\beta\propto 1/\vdisk$). In contrast to Op08, however, we scale the
wind speed so that the total wind energy (per stellar mass formed) is independent of halo mass.
(The total of the wind momentum in Op08 exceeds the momentum available from
photon by almost an order of magnitude). 
The effect of using this `momentum scaling' ($\ahot=1$) in the GALFORM
model is shown in Fig.~\ref{fig:momdriven}. The first panel shows a
model with relatively modest wind speed normalisation
($\vw200=200\kms$), considering a range of mass loading
normalisations $\b200=2-32$. Note that in low mass haloes, the mass
loading will be higher and the wind speed lower. The panel shows that
normalisation of the SMF steadily decreases as the
wind speed increases. The overall SMF is flatter
than that seen in the SW case (where the wind properties are
independent of halo mass, Fig.~\ref{fig:wind_ewind}), and we see that
some models compare favourably with the data above 
a stellar mass of $10^9 \hMsol$.

\begin{figure}
\begin{center} 
\includegraphics[scale=0.4,angle=0]{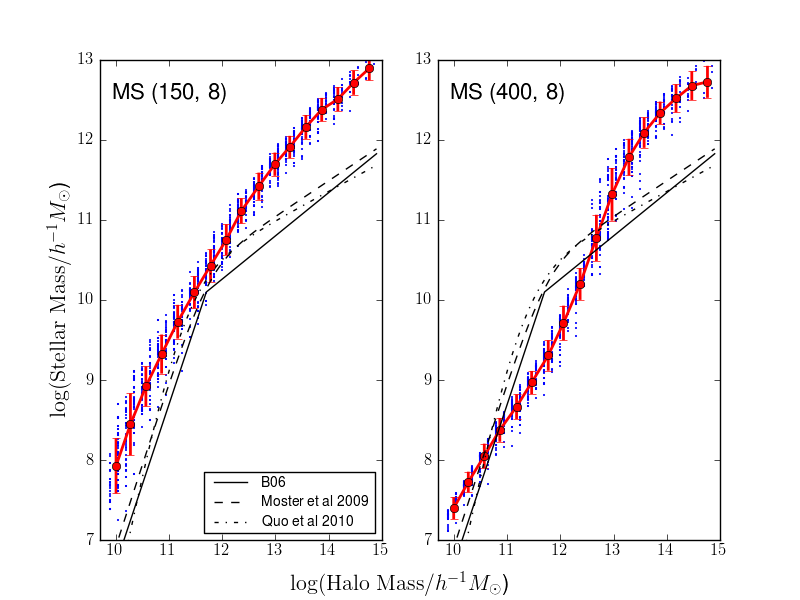}
\end{center}
\caption{Comparison of the $M_*$ -- $\Mhalo$ relation for two momentum scaling models ($\ahot=1$). The first panel shows $(\vw200,\b200) = (150,8)$, the second panel $(400,8)$. Comparing this plot with Fig.~\ref{fig:mstar_mhalo_fixw} shows that the principle effect of the halo mass dependence of the wind is to smooth out the kink in the 
$M_*$ -- $\Mhalo$ relation. See Fig.~\ref{fig:mstar_mhalo_bow06} for
explanation of lines and symbols.
}
\label{fig:mstar_mhalo_mom}
\end{figure}

The second panel shows the effect of increasing the wind speed at a
fixed mass loading, $\b200=8$. As the wind speed increases, the
abundance of $M_*$ galaxies is suppressed. {\rgb As we have seen in the SW
models, a high mass loading can create a dip in the mass function
(where $\vwind\sim\vhalo$). The dip tends to be more smeared out for the MS wind, however.}
The origin of this feature is shown clearly in
Fig~\ref{fig:mstar_mhalo_mom}, where the $M_*$ -- $\Mhalo$ relation is
shown for an MS models with parameters $(\vw200,\b200) = (150,8)$ and
$(400,8)$. The effect of the MS feedback scheme is to introduce a kink
into this relation, with the location of the kink depending on the
wind speed normalisation. The effect is similar to that seen
previously in the SW models, but the kink is more diffuse, resulting in a smoother transition between the high and low mass regimes. The second panel of this figure should be compared with the last panel of Fig~\ref{fig:mstar_mhalo_fixw} as the feedback is the same in $200\kms$ haloes in both cases.
The kink in the  $M_*$ -- $\Mhalo$ relation is clearly much smoother
in momentum driven model, resulting in a less prominent dip in the
mass function. Below $M_*=10^9\hMsol$, the slope of the $M_*$ --
$\Mhalo$ relation is again relatively shallow leading to an over
abundance of low mass galaxies.

{\rgb
The difference between the MS and SW models is also seen in the behaviour of the high mass end of the SMF. At high masses, little of the
feedback material is able to escape the halo, but the mass loading
of the two models differ. As a result, for models with equal normalisation,
the stellar mass associated with high mass haloes in higher in the MS model than in the SW model (compare the last panels of Fig.~\ref{fig:mstar_mhalo_mom} and \ref{fig:mstar_mhalo_fixw}, for example). This is reflected in a greater abundance of high mass galaxies
in the MS model vs.\ SW (compare Fig.~\ref{fig:momdriven} and  \ref{fig:wind_ewind}).  We will show, later, that this has important
consequences for the abundance of the massive galaxies at high redshift.
}

{\rgb
In summary, the dependencies on mass loading and wind speed in the MS model show similar trends to the SW
model. However, the rise in the abundance of the faintest galaxies is
shallower, and the dip in the SMF tends to be smoothed out. With suitable choice of
wind parameters, this model is able to match the observed stellar mass function over a greater range of galaxy mass.
}

\subsection{Energy Scaling Models}

\begin{figure*} 
\begin{center} 
\begin{tabular}{cc} 
\includegraphics[scale=0.4]{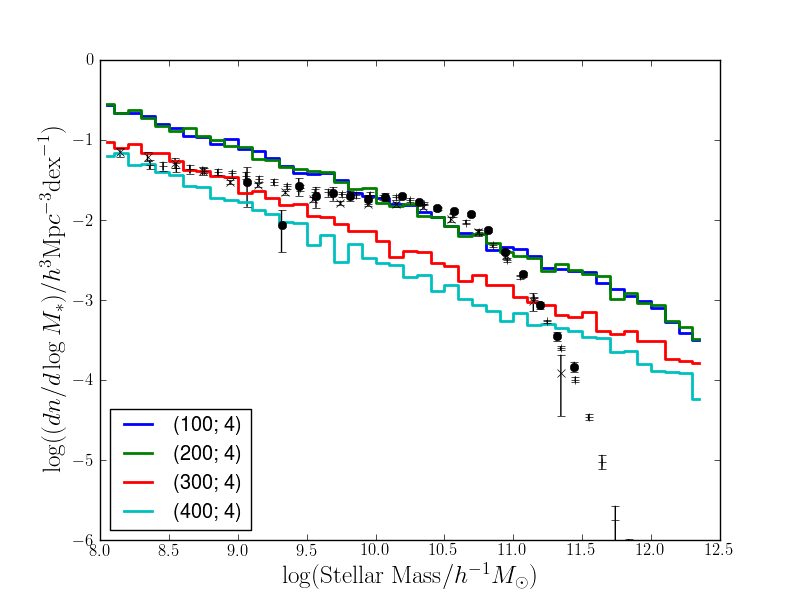}& 
\includegraphics[scale=0.4]{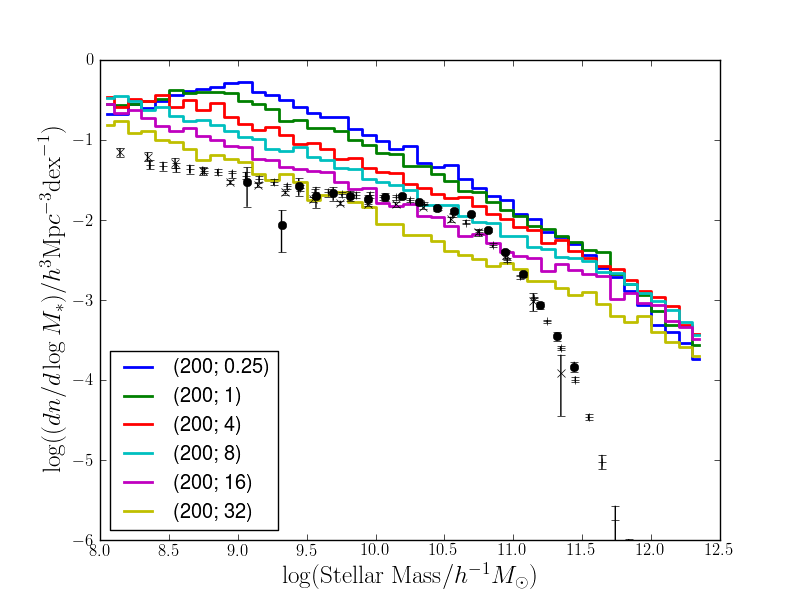}\\ 
\end{tabular} 
\end{center} 
\caption{(left panel): The effect of varying the wind speed in the energy scaling (ES) model ($\ahot=2$).  These models have increasing wind speed normalisation, $\vw200 = 100, 200, 300, 400$; all the models shown have $\b200=4$. Increasing the wind speed normalisation results in further suppression of the SMF. However, the effect is strongest around the knee of the mass function, and the resulting function does not match the observational data.  (right panel): The effect of varying the wind loading at a fixed wind speed.  For this model, the effects of varying mass loading and wind speed are quite similar and 
the mass function cannot be shaped to match the observational data. } 
\label{fig:fountain_wspeed} 
\end{figure*}

Finally, we consider models in which the wind speed is a fixed multiple of the halo circular velocity, such that a fixed fraction of the wind escapes regardless of the halo mass. In the left hand panel of Fig.~\ref{fig:fountain_wspeed}, we fix $\b200=4$ and allow the velocity of the wind to increase, $\vw200=100$ -- $400$. The results for normalised wind speeds between 100 and 200 (blue and green lines) are identical because little material has sufficient specific energy to escape the halo. Further increases in wind speed change the normalisation dramatically as material leaves the halo and takes longer to become available for cooling again. However, the shape of the SMF changes little, and it is not possible to recreate the dip in the mass function that was seen in previous models. In the right hand panel, we show the effect of varying the mass loading of the wind. For this model, the effect is similar to that of varying the wind speed. Since there is no characteristic mass at which the wind stalls, the loss of material from the halo is similar regardless of whether a relatively small fraction of baryons are expelled with high specific energy (and thus remain outside of the halo for an extended period), or a large fraction of material is expelled with lower specific energy.  Finally, we note that the abundance of high
mass galaxies continues the increasing trend seen Figs.~\ref{fig:momdriven} and  \ref{fig:wind_ewind} due to the decrease in mass loading in high
mass haloes (for equal normalisation).

\begin{figure} 
\begin{center} 
\includegraphics[scale=0.4,angle=0]{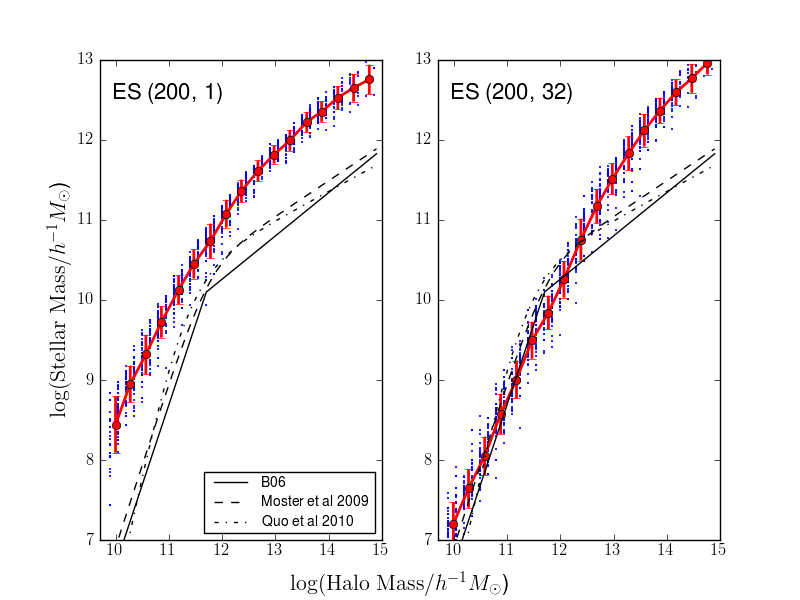} 
\end{center} 
\caption{Comparison of $M_*$ vs.\ $\Mhalo$ for a model with energy scaling of the wind (ES, $\ahot=2$). The two panels show models with $(\vw200,\b200) = (200,1)$ and $(200,32)$. These span the range of models shown in the right hand panel of Fig.~\ref{fig:fountain_wspeed}. See Fig.~\ref{fig:mstar_mhalo_bow06} for
explanation of lines and symbols.} \label{fig:mstar_mhalo_fountain}
\end{figure}

Figure~\ref{fig:mstar_mhalo_fountain} shows the effect of this type of feedback on the $\Mstar$--$\Mhalo$ relation. As can be seen, the kink that enabled us to produce a flat component of the mass function in the first two feedback schemes is absent. The scaling of the wind speed with halo mass means that winds do not stall at a particular halo mass. Instead, the feedback steepens the overall slope of the $\Mstar$--$\Mhalo$ relation. However, although it is closer to the observed relation, the difference in slope leads to a significant mismatch with the SMF, as shown in Fig~\ref{fig:fountain_wspeed}. This clearly illustrates the way in which the normalisation of the mass function is strongly dependent on the slope of the $\Mstar$--$\Mhalo$ relation as well as its normalisation.

Overall, the effect of this type of feedback is less encouraging, and we do not consider this model
further.  The effects of mass loading and wind speed are very similar, and the primary effect of both is to suppress the normalisation of the stellar mass function rather than to alter its shape. While including AGN feedback induces a break at the bright end of the mass function by suppressing cooling in hydrostatic haloes, the faint end slope is not affected. In contrast, the B8W7 model achieves a much improved match to the mass function by adopting a 
stronger halo mass dependence of the wind mass loading.

\section{The Role of AGN feedback}

\subsection{The hot-halo (or ``radio'') mode} \label{sec:radio_mode}

\begin{figure*} 
\begin{center} 
\begin{tabular}{cc} 
\includegraphics[scale=0.4,angle=0]{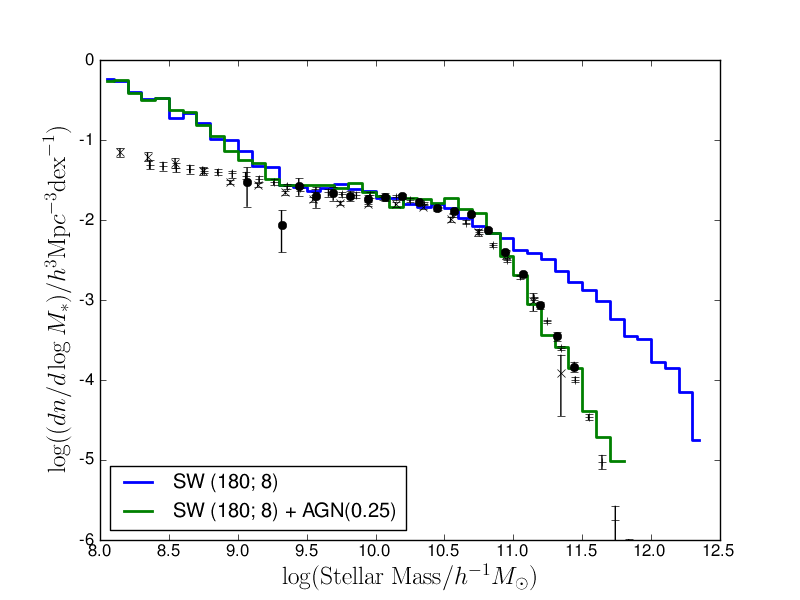} & 
\includegraphics[scale=0.4,angle=0]{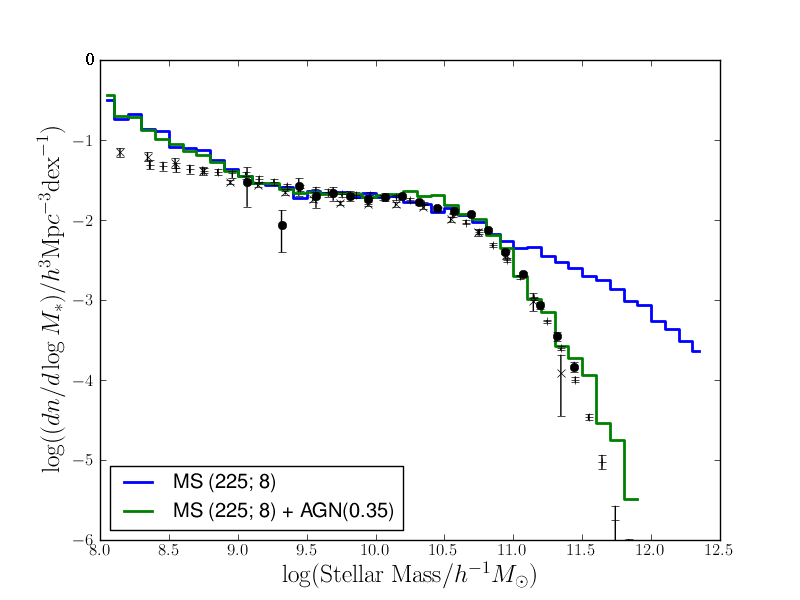} \\ 
\end{tabular} 
\end{center} 
\caption{(left panel): Effect of ``hot-halo'' AGN feedback on the SW model. Both models have $(\vw200,\b200) = (180, 8)$, and we compare $\acool=0.0$ and 0.25. Increasing $\acool$ adjusts the ratio of cooling and free-fall times at which AGN are assumed to become effective. The increase shifts the break in the mass function to lower stellar mass so that the model matches the break in the mass function well.  (right panel): We show the effect of AGN feedback in the MS ($\beta\propto 1/\vdisk$) model. In this case, we compare models with $(\vw200,\b200) = (225, 8)$ for $\acool=0.0, 0.35$. In both panels, black points show observational data. 
We have supplemented the \citet{bell2003} (circles) and
\citet{li_white09} (pluses) with preliminary data from the GAMA survey (crosses, Baldry et al., in prep)
to provide an independent assessment of the observational uncertainties at $\Mstar < 10^9\hMsol$. 
The up-turn in the model mass functions at low $\Mstar$ is inconsistent with both recent data-sets.}
\label{fig:with_agn}
\end{figure*}

None of the models discussed so far are able to match the abrupt turn over of the SMF. In this section, we consider the ``hot-halo'' mode of AGN feedback, associated with the heating of the group and cluster diffuse material. Bow06 argued that the AGN feedback loop can only be established if the 
cooling time is longer than the dynamical time (or sound crossing time) of the halo, and showed that allowing
AGN to suppress cooling in hydrostatic haloes resulted in a good description of many galaxy formation
properties.  Bow08 extended this by allowing the AGN to expel material from hydrostatic haloes (rather than simply replacing the energy radiated). They showed that this model was able to match the observed X-ray scaling relations of groups and clusters as well as many of the observed properties of galaxies.  Typically, the heat input is assumed to be associated with low-excitation radio sources \citep{croton2006, best2007}, however, the exact heating mechanism is not important to the scheme. The crucial distinction is that only hydrostatic haloes are affected and that the cold gas
disk of the galaxy is affected only indirectly because of the reduction in the supply of cooling gas (\citet{vandevoort2011} discuss these effects
in the context of hydrodynamic simulations).
{\rgb
It is important to note that the effectiveness of AGN feedback is not a threshold imposed at a fixed mass, but is the result of dynamically tracking the relative cooling and dynamical times of the halo as it evolves.
}

The results of including this type of AGN feedback in the SW and MS model are shown in Fig~\ref{fig:with_agn}. In each model we choose the wind speed and mass loading to achieve a good match to the abundance of galaxies with stellar mass between $10^{9.5}$ and $10^{11}\hMsol$ and have then adjusted the $\acool$ parameter to achieve a good match to the observed stellar mass function. Increasing $\acool$ adjusts the ratio of cooling and free-fall times at which the AGN is assumed to become effective. Larger values make AGN feedback more effective and shift the break in the mass function to lower stellar mass.  With a suitable value for $\acool$ the models match the observed mass function well. 
For the SW model, we find that $(\vw200,\b200) = (180, 8)$ gives a reasonable match to the observed SMF if combined with $\acool= 0.25$.  For the MS model, we find that $(\vw200,\b200) = (225, 8)$ combined with $\acool=0.35$ gives a good description of the observational data for stellar masses
above $\sim 10^9\hMsol$. Both models over predict the abundance of the lowest mass galaxies, although this
problem is reduced in the MS model.  To emphasise this point, we have superposed pre-liminary data 
from the GAMA survey (Baldry et al., in prep). This provides an independent assessment of the 
mass function for low mass galaxies. Although the detailed shape of the mass function differs slightly
from \citet{li_white09}, the differences are much smaller than the discrepancy between the models
and the observational data.  In contrast the B8W7 model keeps a shallow mass function slope down to the faintest
galaxies plotted (see Fig.~\ref{fig:Bow06}). {\rgb The difference in
behaviour arises from the steeper slope of the stellar mass -- halo mass
relation in B8W7.} 

\begin{figure*} 
\begin{center} 
\includegraphics[scale=0.8,angle=0]{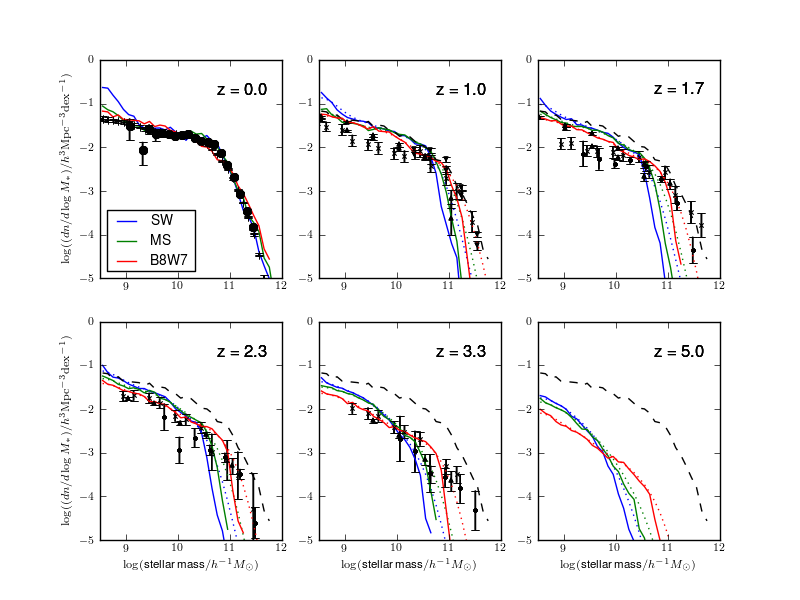} 
\end{center} 
\caption{The evolution of the SMF compared between three models that match the $z=0$ SMF well: B8W7 (red), SW (blue), MS (green). All three models include AGN feedback, and the $z=0$ stellar mass function of B8W7 is repeated as a black dashed line in each panel.  {\rgb The effect of observational errors are indicated by the dotted lines which show the
effect of convolving the models with a stellar mass error of 0.2 dex. 
For comparison, we show observational data for the high redshift mass function from a variety of sources (see text for details) as black symbols. Out to $z=2$, all three models result in similar evolution of the stellar mass function. The break in the mass function evolves more quickly in the SW and MS models than in the B8W7. However, the
deficit of massive galaxies in the MS model could be accounted for if the 
random stellar mass errors are greater than 0.2 dex. All of the models
show an excess of $\Mstar\sim10^{10}\hMsol$ galaxies at $z=1-2$ compared
to most observational data sets.
}
} 
\label{fig:compare_starmf_evo} 
\end{figure*}

{\rgb
In Fig.~\ref{fig:compare_starmf_evo}, we compare the evolution of the mass function for the models discussed above. The figure also shows the B8W7 model, and recent observational data  from \citet{drory2005,
bundy2005, marchesini2009} and \citet{mortlock2011} (plus, circle,
cross and triangle respectively). All of the models include AGN feedback following the Bow08 scheme.  The key issue that we wish to test with this plot is whether the models generate sufficient massive galaxies at high redshifts, and we focus on the brightest galaxies at each epoch.
Compared with B8W7 and the observational data, the new models show a rapid
decrease in the abundance of the most massive galaxies at higher redshift.  The discrepancy is worst for the SW model. 

The differences in the behaviour of the models can be traced to the differences in wind mass loading in high mass haloes (see \S\ref{sec:effect_of_wind_params}). The effect arises since the ``hot-halo'' mode of feedback is {\it not} a simple cut-off in cooling at high halo
mass, but explicitly compares the halo cooling time and dynamical time,
taking into account the halo formation history. In practice the effective
halo mass threshold increases slowly with redshift. Moreover, since none of the models
can eject material from massive haloes, the efficiency of star formation
is inversely proportional to the mass loading. The net result is that
massive galaxies appear at higher redshifts in the model with the strongest halo mass dependence of the mass loading. Since $\beta \propto \vhalo^{-3.2}$ in the B8W7 model, this model provides the
best match to the observational data, followed by the MS model ($\beta \propto \vhalo^{-1}$). 
}

We should,
however, note that there are considerable random uncertainties in determining
the stellar masses of high redshift galaxies. Applying this convolution will tend to smear the model predictions, resulting in a tail of higher mass galaxies (see discussion in \citet{marchesini2009}).  Thus while this data-set set picks out the B8W7 model, careful analysis of the 
observational errors is required before reaching a definitive
conclusion. {\rgb In order to illustrate the effect of random
  uncertainties in the mass determination, dotted lines show result of
  convolving the model with a random error of 0.2 dex. This has a
  pronounced effect on the abundance of the most massive galaxies, as
  a small population of galaxies that are mistakenly assigned low
  stellar mass can easily overwhelm the true population. The
  comparison still favours the B8W7 model, but assigning larger mass
  errors would make it difficult to exclude the MS model with high confidence.}

{\rgb  Focussing on lower
mass galaxies, we see that all of the models appear to over-predict the observed normalisation of the mass function at $z=1$--2 \citep{marchesini2009}. Although there is considerable scatter between data-sets,
and the survey volumes are relatively small, this does appear to be a
persistent problem, and only the data from \citet{drory2005} are
consistent with the evolution seen in the models. This discrepancy is
also evident if the K-band luminosity functions are compared directly
(eg., \citet{cirasuolo2010}). \citet{pozzetti2010}
suggest that the problem lies with the mass dependence of the specific
star formation rates of the model galaxies, and we will examine this
in Section~\ref{sec:ssfr}. Our preferred interpretation is that the data
require a stronger redshift dependence of feedback. We have already
shown that the pGIMIC model provides a good description of the mass
function at $z>1$, so a promising route would be to vary the wind
speed parameter between 180 at $z=0$ and 275 at $z=1$. Alternatively,
the required variation in the fraction of the wind escaping
would naturally arise if we were to choose a criterion for
wind escape based on the halo mass rather than circular velocity, at
least at low redshift. It is unclear why this choice should be
physically motivated, however.  Perhaps a better explanation could be 
the greater gas content of high redshift disks, and thus the tendency
for star formation to occur in more massive star forming complexes
\citep{jones2010, genzel2011}.}

{\rgb
In summary, introducing a hot halo mode of feedback creates a break in the stellar mass 
function in all three models. As a result, all three provide a good match to 
the observed SMF above a stellar mass of $\sim 10^{9.5} \hMsol$.
At lower stellar masses, the mass function of the SW model rises steeply, and is 
inconsistent with the observational data. This trend is less pronounced in the MS model,
while the B8W7 model has a flat SMF to much lower masses.  The models
predict different evolution of the SMF, with B8W7 showing the highest 
abundance of massive galaxies at $z=1$ and above. All three models predict an abundance
of $10^{10} \hMsol$ galaxies at $z=1$ that appears to be at odds with
the data, and suggest that the effective wind speed should be higher
at $z>1$ than at the present day.
}

\subsection{The ``Starburst'' (or ``QSO'') mode'}

\begin{figure} 
\begin{center} 
\includegraphics[scale=0.4,angle=0]{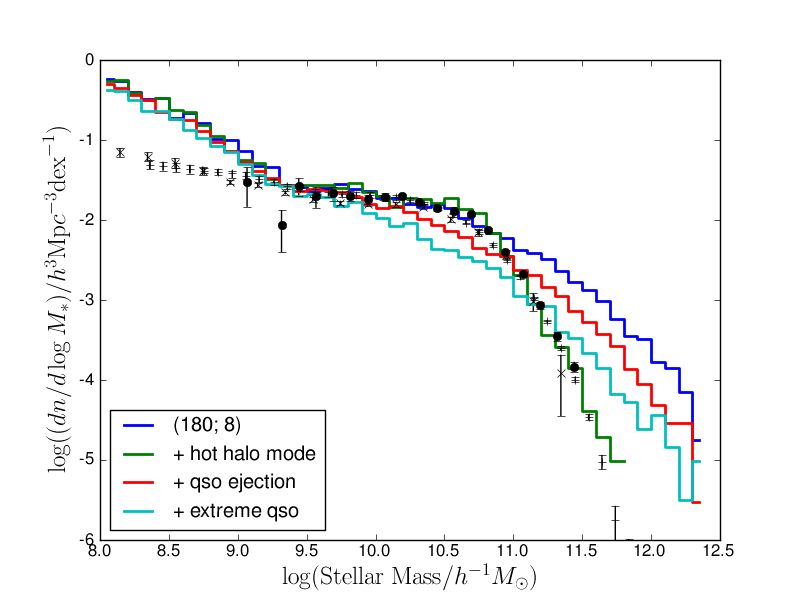}
\end{center} 
\caption{(Left Panel) Starting from the SW model [$(\vw200,\b200) = (180, 8)$, blue line], we contrast the effect of the AGN ``hot-halo'' mode feedback scheme (implemented by setting $\acool =0.25$, green line) with the effect of inducing a wind with very high mass loading ($\b200=16$) and wind speed ($\vw200=1130\kms$) during bursts of star formation (red line). We assume that the energy to drive such powerful outflows comes from the QSO phase of AGN growth. In order to show the effect of more frequent outbursts, we also show an ``extreme QSO'' model in which disks are much more unstable than in B8W7
(cyan line), so that black hole fueling occurs 
more frequently. The main effect of the QSO mode is to suppress the formation of galaxies near the break of the mass function and to drive the mass function towards a power-law shape. The sharp break in the mass function is only created if AGN feedback is only effective in hydrostatic haloes. 
}
\label{fig:agn_or_qso} 
\end{figure}

{\rgb
Another channel of AGN feedback, often referred to as the ``QSO'' or ``starburst'' mode, also
has the potential to be important because a large fraction of the black
hole mass is accreted in this way. The GALFORM model assumes that black hole growth is triggered
when gas is transported to the center of a galaxy by disk instabilities or galaxy mergers.
Most of the cold gas fuels a burst of star formation, but a small fraction is 
accreted by the black hole (eg., \citet{springel2005}).  Since most mergers are gas rich, this results in a strong correlation between black hole mass and bulge mass very like that observed. We will use the term ``QSO mode'' and ``starburst mode'' interchangeably, perhaps the term ``starburst'' should be preferred since it makes it clear that this channel only occurs during such events.  The key distinction is that the ``starburst'' mode acts
on the cold gas of the host galaxy, rather than acting through the
heating of hot gas in the haloes of galaxy groups and clusters.  In
the starburst mode feedback may lead to explosive winds that blow cold gas out
of the host galaxy.  If cold gas is removed from the system at
sufficiently high specific energy it will suffer a long delay before
it is able to cool once again. This type of feedback has been explored
in idealised numerical simulations which have shown that the
energetics of the black hole can plausibly remove the whole
interstellar medium of the merging galaxies \citep{springel2005,
  hopkins_hernquist2006} (although higher resolution simulations suggest that the geometry of the central outflow may play an important role, \citet{hopkins2010}).}
However, this channel expels only the cold material from the system, and does not prevent further accretion. As the halo grows, it accretes new satellite galaxies, together with their gas, so that (in practice) star formation quickly re-establishes itself.

Fig.~\ref{fig:agn_or_qso} illustrates the effect of the QSO mode of expulsion.  All of the models we
consider reproduce the observed correlation between the mass of the black hole and the mass of the 
galaxy bulge. We start from the SW model (with parameters $(\vw200,\b200) = (180, 8)$, blue line). In this model,
black holes grow strongly as a result of galaxy mergers and disk instabilities (see Bow06), but 
this results in no effective feedback. The default model assumes that all the energy generated
in black hole events is radiated without doing significant mechanical work. In order to explore
what would happen if this radiation coupled effectively to the surrounding gas (or if the quasar accretion disk
produced a high speed wind), we implement a ``QSO mode'' of feedback by using much stronger feedback during starbursts (compared to quiescent star formation events). We illustrate the effect by using $(\vw200,\b200) = (1130, 16)$ during bursts (the results are similar for other parameter choices) so that the energy of the wind during the burst is 80 times larger than that during quiescent star formation. The figure shows that even winds of this strength have a modest effect on the mass function. Furthermore, their effect is to suppress the abundance of $M_*$ galaxies rather than to create an exponential break in the SMF.  We can
produce a stronger effect on the mass function by increasing the frequency of starbursts. A simple
way to achieve this is to tighten the disk stability criterion so that disks more frequently become
unstable.  The effect is illustrated by the ``extreme QSO'' model in the plot (cyan line).  The model has been 
shifted further from the observed SMF, giving the mass function an almost power-law form.

{\rgb
We can compare these models with \citet{gabor2011} who modify the hydrodynamical models of
Op08 to investigate the effect of quenching star formation after galaxy mergers.  They
contrast this form of feedback with a model in which star formation is suppressed in
hot haloes. The implementation of their schemes are similar to those adopted here
(although their quasar-mode feedback scheme is triggered only by mergers, while
it is triggered by both mergers and disk instabilities in our model), and
the results are qualitatively similar.  In particular, the merger model tends to have a 
relatively weak effect on the overall mass function, and fails to imprint a characteristic
scale on the SMF. 
}

We have experimented with using other models as a starting points.
If we start from a model with much weaker quiescent feedback, a strong ``starburst'' mode feedback fails to reproduce the shape of the observed SMF, again tending to drive the mass function towards a power-law. The ``starburst'' mode does not have the required effect because it does not scale strongly with halo mass (as is the case for the ``hot-halo'' mode). In summary, while the starburst/QSO channel might supplement the feedback from supernova during star bursts, it does not provide a scheme for creating a break in the stellar
mass function.

\section{Further Considerations}

\subsection{The Star Forming Sequence}
\label{sec:ssfr}

\begin{figure}
\begin{center}
\begin{tabular}{c}
\includegraphics[scale=0.4,angle=0]{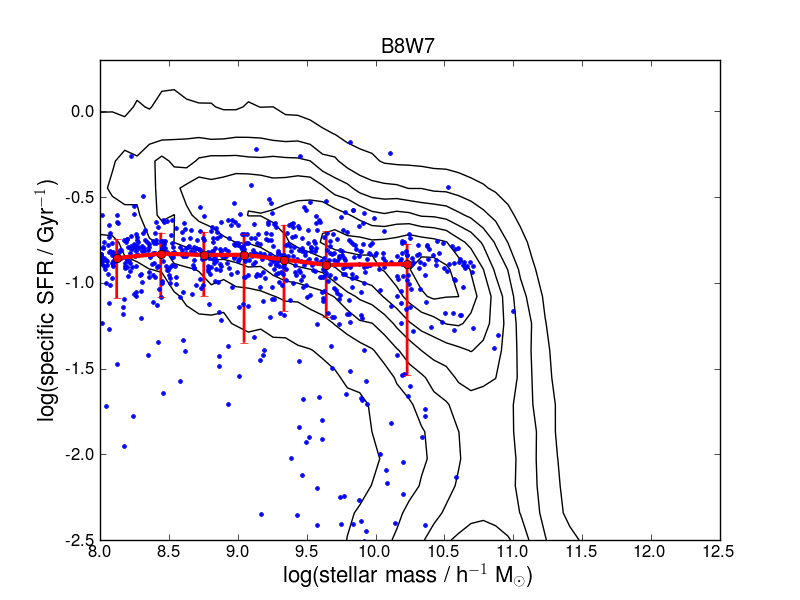}
\\
\includegraphics[scale=0.4,angle=0]{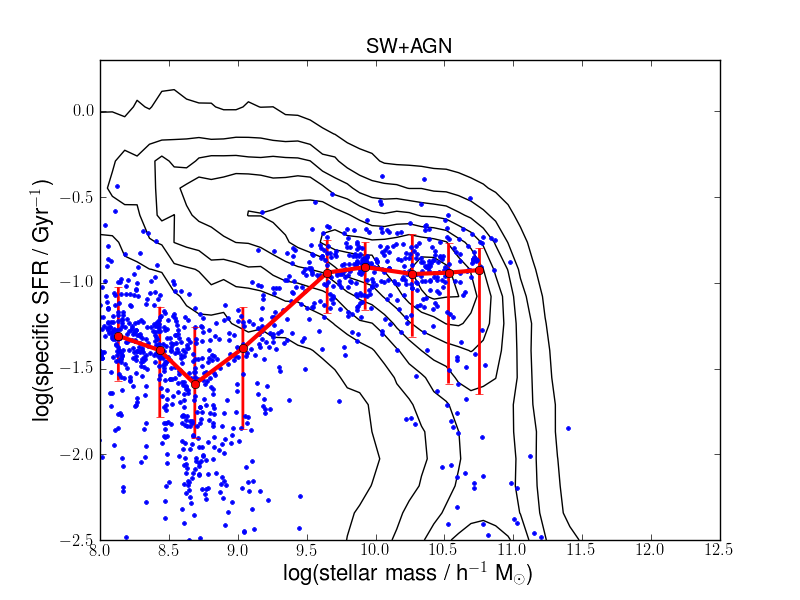}
\\
\includegraphics[scale=0.4,angle=0]{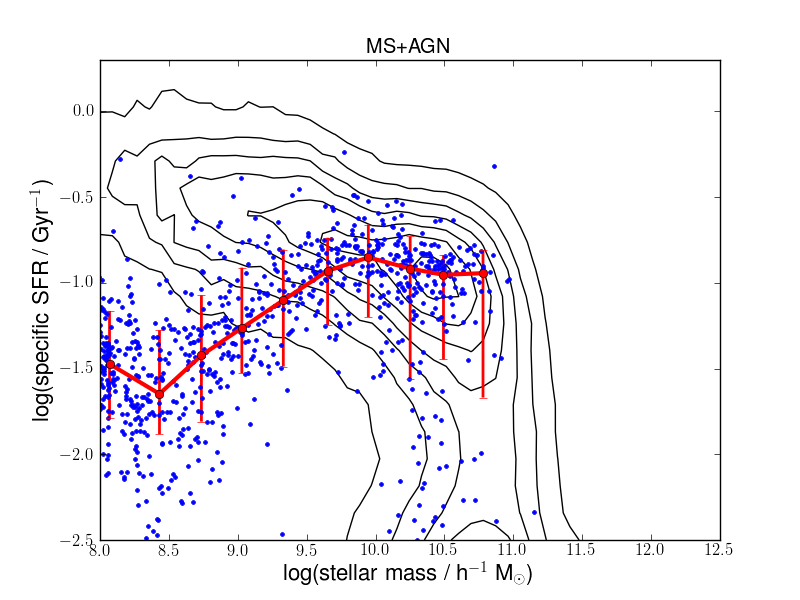}
\\
\end{tabular}
\end{center}
\caption{Comparison of the specific star formation rates of central galaxies for three models
which provide encouraging fits to the SMF. In seperate panels, we show the baseline model,
B8W7, SW+AGN [$(\vw200,\b200,\acool) = (180,8,0.25)$] and the
MS+AGN model [$(\vw200,\b200,\acool) = (250,8,0.35)$]. The star formation sequence for local galaxies is shown as black contours (see text for details). A random sample of model galaxies
are shown as blue points, with the median relation and 10\% and 90\% percentiles
shown as a red line with error bars. The B8W7 model has a constant specific 
star formation rate regardless of stellar mass.
In contrast, the SW and MS models produce relations that are strongly dependent on system mass
and are incompatible with the observational data: instead of a near constant specific star formation 
rate, the SW and MS model have specific star formation rates that are lower in small galaxies.  
}
\label{fig:ssfr_agn}
\end{figure}

We have compared the different schemes on the basis of the stellar mass function and the central stellar mass. In this section, we compare the models with the star formation rates of the central galaxies. We focus on B8W7 and the best fitting models with constant wind and $\beta\propto 1/\vdisk$ feedback scalings, SW+AGN [$(\vw200,\b200,\acool) = (180,8,0.25)$] and MS+AGN  [$(\vw200,\b200,\acool) = (225,8,0.35)$].  The SMFs
of these models are shown in Fig.~\ref{fig:with_agn}.

Fig.~\ref{fig:ssfr_agn} shows the logarithm of the $z=0$ specific star formation rate (SSFR) as a function of stellar mass. We focus on the properties of star forming galaxies (which we define as having SSFR $>0.01 \Gyr^{-1}$).  In the model we restrict attention to central galaxies to avoid uncertainties in the treatment of satellite galaxies. In all the models, there is a clear sequence of star forming galaxies that can be cleanly compared to the observed star forming sequence.  There is also a large population of galaxies which are not seen on this diagram because their star formation rates are extremely low. These are satellite galaxies, or central galaxies in haloes with effective AGN feedback.  

We compare the theoretical models to observational data from \citet[][updated to DR7]{brinchmann2004}. The observational data is shown as black contour lines in the figure. The observed relation is almost flat (with low mass galaxies having slightly higher SSFR) up to stellar masses of $10^{11}\hMsol$ (above which star formation is suppressed by AGN feedback). Given the age of the universe, the location of the sequence at SSFR $\sim 10^{-0.8} \Gyr^{-1}$ implies that the specific star formation rates of all galaxies have been steady (or slowly rising) over the history of the universe. Note that the uncertainties due to dust obscuration would tend to increase the star formation rates of the most massive galaxies.  However, the data presented already include an extinction correction based on the Balmer decrement, and agree well with specific star formation rates based on SED fitting \citep{mcgee2011}. 

In each panel, a random sample of model galaxies are shown as blue points, while the red line with error bars shows the median specific star formation rate and the 10$^{\rm th}$ and 90$^{\rm th}$ percentiles of the distribution. We include only central star forming galaxies (with SSFR $>0.01 \Gyr^{-1}$) in this calculation, but very similar results are obtained if we include star forming satellite galaxies as well. The observed relationship is reproduced fairly well by B8W7.  The specific star formation rate in the model is flat over a wide range in stellar mass, from below $10^7\hMsol$ to almost $\sim 10^{11}\hMsol$ (where the relation dips as the supply of fuel for star formation is suppressed by the hot-halo feedback). 

In contrast, the SW and MS models predict a relation with a noticeable {\it decline} in SSFR towards
lower stellar masses. Although the declining trend is less abrupt in the MS model (bottom panel) than in SW (middle panel), both relations are clearly inconsistent with the observational data.
We infer from this figure that a dramatic change of feedback efficiency cannot be responsible for the flattening of the SMF. This problem is also seen in the hydrodynamical models of GIMIC and Op08 (see \citet{dave2011}), although the limited mass resolution of those simulations limits the comparison to galaxies more massive than $\sim 10^9\hMsol$, and consequently the mass dependence is not so clearly evident.

{\rgb
It is interesting to understand the origin of the dip. In the SW and
MS models, the flat region of the SMF is created by a transition
between the two feedback regimes: for low halo masses, feedback is
extremely effective at expelling gas from the halo and the star
formation rate is strongly suppressed. At higher masses, however, the
wind velocity is no longer sufficient to escape the halo and the cold
gas mass and star formation rate increase. However, because the division 
between the two regimes occurs at a fixed escape velocity, we expect that the
halo mass of the transition evolves rapidly with redshift, $M_{\rm halo, kink} \sim (1+z)^{-3/2}$. Allowing for the dependence of stellar mass on halo mass,  $\Mstar\propto\Mhalo^2$ (eg. Fig.~\ref{fig:mstar_mhalo_bow06}, the relation evolves slowly with redshift) we expect the stellar mass at which the transition occurs to evolve as $M_{\rm *, kink} \sim (1+z)^{-3}$. Thus the
transition mass evolves more quickly
than the mass of an individual galaxy. Thus the transition mass is much smaller at high redshift. Over time, an individual galaxy makes a transition
from the regime in which ejection is in-effective to the one in which it is.
Consequently, galaxies somewhat below the transition mass at $z=0$ have low current star formation rates compared to their past average. 
At the very lowest stellar masses, the SSFR in the SW and MS models begins to increase. Galaxies that lie well below the kink in the $\Mstar - \Mhalo$ relation have experienced similar feedback during their formation history and the rise is thus to be expected. 
}

{\rgb
Although the B8W7 model provides the best description of the observational
data, it does not reproduce the weak trend for the SSFR to increase 
as the stellar mass decreases (SSFR $\propto \Mstar^{-0.22}$) that is seen in the data.
The strength of this trend is controversial, but does not appear to be an observational selection effect. It is seen regardless of the star formation diagnostic that is applied, and is apparent at higher redshifts as well as locally (but this depends critically on the definition of the sample --- for a recent overview, see \citet{karim2011}). In \S\ref{sec:radio_mode},
we noted that the surprisingly rapid evolution of the normalisation of
the observed mass function suggested that the effective wind speed should
scale with redshift. This change would also have implications for the SSFR, since the present-day star formation rate would increase relative to the past average. Varying the escape speed rather than the mass loading
could create a tilt in the SSFR $- \Mstar$ relation since the effect
will be strongest around the kink in the $\Mstar - \Mhalo$ relation
but result in little change in the star formation histories of the 
most massive star forming galaxies.
}

{\rgb
In summary, the specific star formation rate of galaxies provides an important additional discriminant of the models. The B8W7 model comes closest to matching the observed 
data, with the characteristic specific star formation rate that is almost independent
of stellar mass. In contrast the SW and MS models show specific star formation rates that decline with decreasing stellar mass, while the observational data show an slightly increasing trend. Further exploration of feedback schemes that scale systematically with redshift is required to identify a model
which produces a better match to the observational data.
}

\subsection{The Star Formation History of the Universe}

\begin{figure}
\includegraphics[scale=0.4, angle=0]{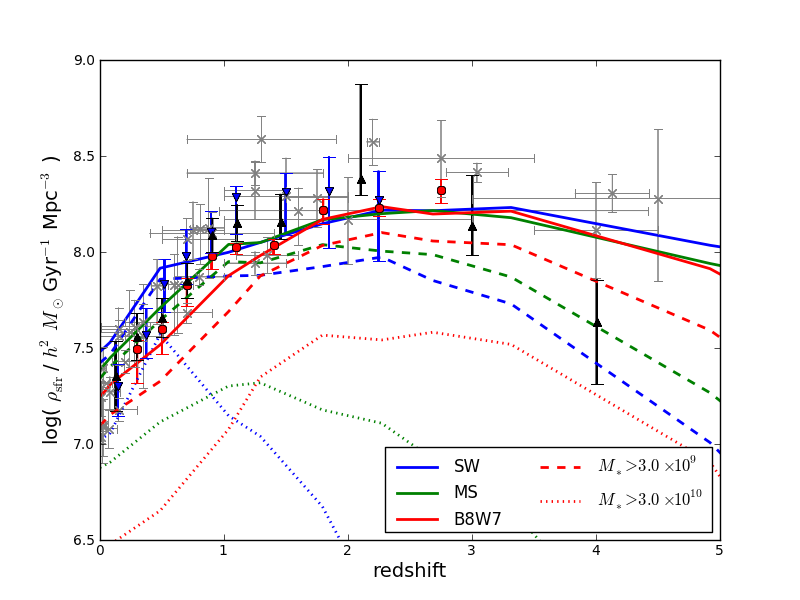}
\caption{{\rgb The evolution of the cosmic star formation rate density 
as a function of redshift. Solid lines show the total star formation rate
density. Line colours distinguish different feedback models.
All three models include hot-halo AGN feedback that suppresses the formation of stars in high mass haloes. While the models all  show similar star formation rates above $z=2$, the decline in the star formation rate between $z=1$ and the present day differs markedly. The solid
lines should be compared with the observational data, shown as black crosses
(from the compilation of \citet{hopkins2004}), red circles (from the 
stacked VLA analysis of \citet{karim2011}), blue triangles from \citet{rodighiero2010} and black triangles from \citet{cucciati2011}.  The dashed and dotted lines show the contribution 
from galaxies more massive that $10^{9.5}$ and $10^{10.5}\hMsol$ respectively. Despite overall the similaraity of the star formation histories in the three models, the way the star formation rate is divided between stellar masses varies greately.
}} 
\label{fig:rhostardot}
\end{figure}

\begin{figure}
\begin{center}
\includegraphics[scale=0.4,angle=0]{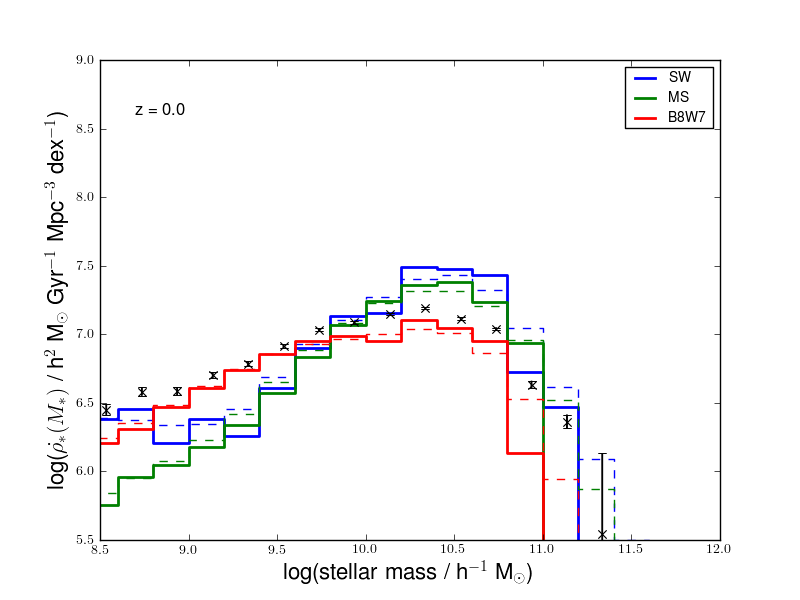}
\includegraphics[scale=0.4,angle=0]{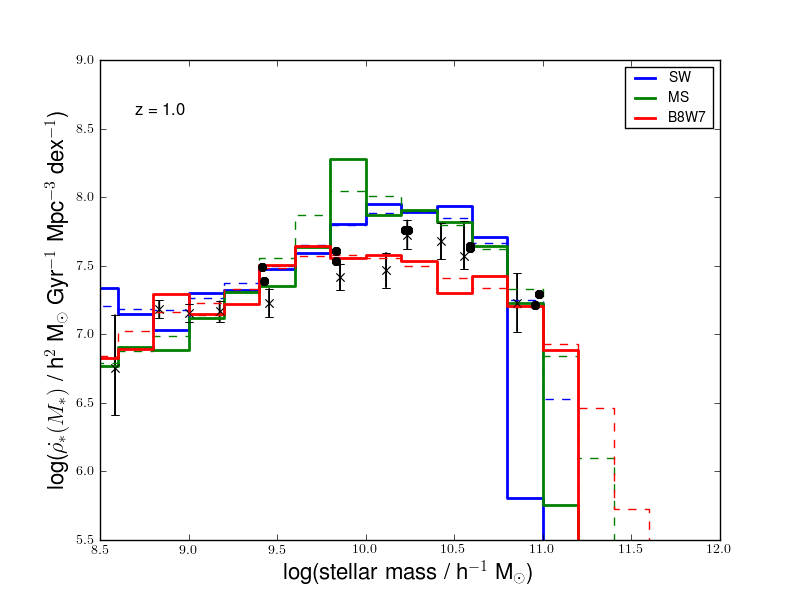}
\includegraphics[scale=0.4,angle=0]{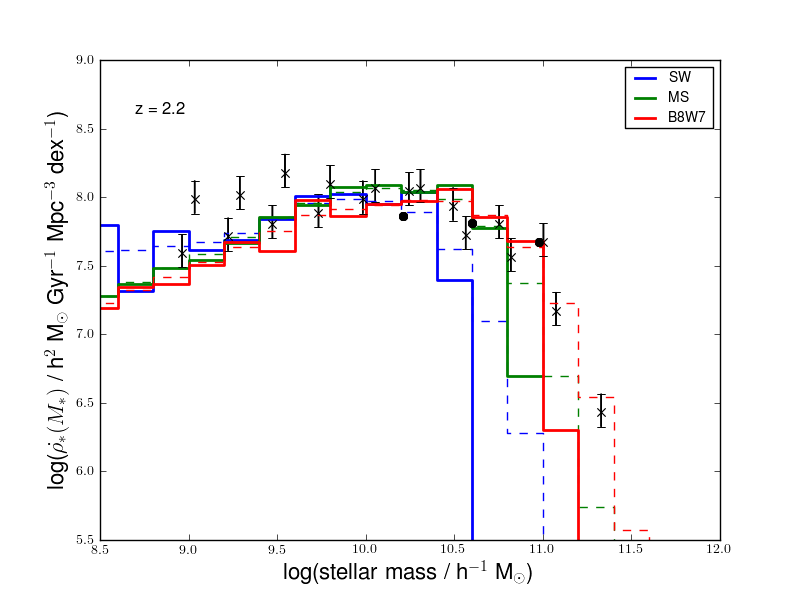}
\end{center}
\caption{
{\rgb
The panels in this figure show the evolution of the cosmic star formation rate density as a function of stellar mass.  The coloured lines differentiate 
different feedback schemes, while the panels compare to data at $z=0.0,
1.0$ and 2.2. Solid lines show the distribution obtained from the
models, dashed lines illustrate the effect of a random error of 0.2 dex
in the stellar mass estimates. Observational data collected by \citet{gilbank2011} and 
\citet{karim2011} are shown as black symbols. This figure gives a more
accurate comparison of the observed star formation rate density with
that predicted by the models (see text for discussion).
}}
\label{fig:rhostardot_evo}
\end{figure}

{\rgb
We have seen that all the models reproduce the observed build-up of stellar mass reasonably well, another way to tackle this question is to investigate 
the star formation rates of galaxies directly.  In Fig.~\ref{fig:rhostardot}
we show the evolution of the cosmic star formation rate. For the models
this is calculated by integrating the contribution of galaxies down to
stellar masses of $10^8\hMsol$.  As a result of including feedback from AGN in 
order to suppress the formation of the most massive galaxies, the behaviour of all three models is broadly similar in this plot. Above $z=2$, the integrated
star formation rates of the models are very similar. However, the strength
of the decline in the star formation rate differs between the models,
being much stronger for the B8W7 model than for the SW scheme.  As expected,
the MS scheme lies in between.

We compare the models with 
the compilation of observational data from \citet{hopkins2004} and the
more recent data of \citet{karim2011} (radio, red circles), \citet{rodighiero2010} (Mid-IR, blue triangles) and \citet{cucciati2011}
(rest-frame UV).  The radio and mid-IR
based measurements have the advantage of that the obscured star formation
is accounted for (see \citet{karim2011} for further discussion).
None of the models match the trends in the observational data perfectly.
Given the scatter in the observational data sets there is little reason
to choose between the MS and B8W7 models on the basis of this plot. However, the uncertainties in the observational data arise in large part because of the large extrapolation required to correct the observed star formation
rate density for galaxies that are too faint to be directly detected.

An important question is therefore to examine how the star formation
rate density depends on the stellar mass of galaxies. The mass dependence
of the cosmic star formation rate in the models is illustrated 
by dashed and dotted lines in Fig~\ref{fig:rhostardot}. The line-styles
show the integrated contribution from galaxies more massive than $10^{9.5}$
and $10^{10.5}\hMsol$ respectively.  Although the total star formation rates of the models are similar, the way the star formation is distributed between 
galaxy masses is different, with the SW and MS models showing a shift from 
star formation that is dominated by low mass ($<10^{9.5}\hMsol$) galaxies at high redshift (note the large difference between the solid
and dashed curves) to a dominance by lower mass galaxies at low redshift.  In contrast, in the B8W7 model massive galaxies make a similar contribution to the total star formation rate density at all redshifts (ie., there is a constant offset between the solid, dashed and dotted lines). Thus, although all three models show a similar total star formation rate density at
$z>2$, the contributions of different mass galaxies is very different in the models. 

Given the different contributions of high and low mass galaxies in the
models, it is very dangerous to quantitatively compare to observational
data for the total star formation rate density since the observations
are usually based on extrapolation of the properties of high mass
galaxies. Therefore in Fig.~\ref{fig:rhostardot_evo} we show the mass
dependence of the cosmic star formation rate, separating the redshift
dependence into separate panels. The different models are
distinguished by the coloured lines. The solid lines show the
distribution measured for the model galaxies, and the dashed lines
illustrate the effect of a 0.2 dex error in stellar mass
assignment. Using this plot, it is then possible to directly compare
to observational measurements based on mass-complete samples
\citep{gilbank2011, karim2011}. Historically, such plots have been use
to infer that the star formation density is dominated by large
galaxies (at high redshifts), and by lower mass galaxies at low
redshift. In practice, the more complete and sufficiently deep data
sets show that the increase in the star formation rate density with
redshift is similar for all galaxy masses, however, the limitations of
the observed data-sets, particularly at high redshift are evident.

Comparison of the models and observational data in this logarithmic plot
allows the contribution from the tails of the mass distribution to be
clearly seen. At low redshift, star formation in the
SW and MS models is more concentrated towards high mass galaxies
than suggested by the data. In
contrast, star formation in the B8W7 model is somewhat too flat. Also, while  a significant contribution to the star formation rate comes from
low mass galaxies (as suggested by the observations), the model
fails to generate sufficient massive star forming galaxies.  This occurs because the AGN feedback parameters need to match the observed SMF set an effective halo mass threshold that is lower in B8W7 than the 
SW and MS models. This is driven by the larger scatter in galaxy mass
at fixed halo mass in B8W7. Thus the model contains galaxies of the stellar mass required to match the observed SMF, but star formation has been suppressed in too many of them. A stellar mass error
of 0.2 dex improves the match (as shown by the dashed line), but this probably
over-estimates the uncertainty in the local data. The difference in
behaviour at $z=0$ tallies with the differences in the models'
SSFR--$\Mstar$ relations seen in Fig.~\ref{fig:ssfr_agn}.

The behaviour of the models is similar at $z=1$: B8W7 results in
a distribution that is flatter than the data, while the MS and SW
models are slightly too peaked. The order of the models at the massive
end is reverse, with the B8W7 model suggesting the greater population
of massive star forming galaxies. The differences are small, however,
and likely to be masked by the uncertainties in the stellar mass
determination (see dashed lines).  At $z=2$, a deficit of 
massive star forming galaxies is apparent for the SW model, and the rapid
drop at high masses is not reconciled with the observational data
if an error 0.2 dex is assigned to the stellar mass. This deficit is 
to be expected from our comparison with the SMF of
this model: the SW model lacks sufficient high mass galaxies.

In summary, while all models result in similar predictions for the 
evolution of the integrated cosmic star formation rate, the mass 
dependence of the cosmic star formation rate exposes important
differences between the models. Although none of the
trends are conclusive, the comparison highlights different issues
with the models. The SW and MS models tend to under-represent star formation
from low mass galaxies at low redshift. At high redshift, the 
SW model under-represents the contribution to star formation from high
mass galaxies. In contrast, the B8W7 tends to under-represent the contribution
from high mass galaxies at lower redshift. As the consistency
and completeness of the observational data improves this approach 
has great potential for discriminating between feedback models. 
}

\subsection{Where are the baryons?} 
\begin{figure}
\begin{center}
\begin{tabular}{c}
\includegraphics[scale=0.35,angle=0]{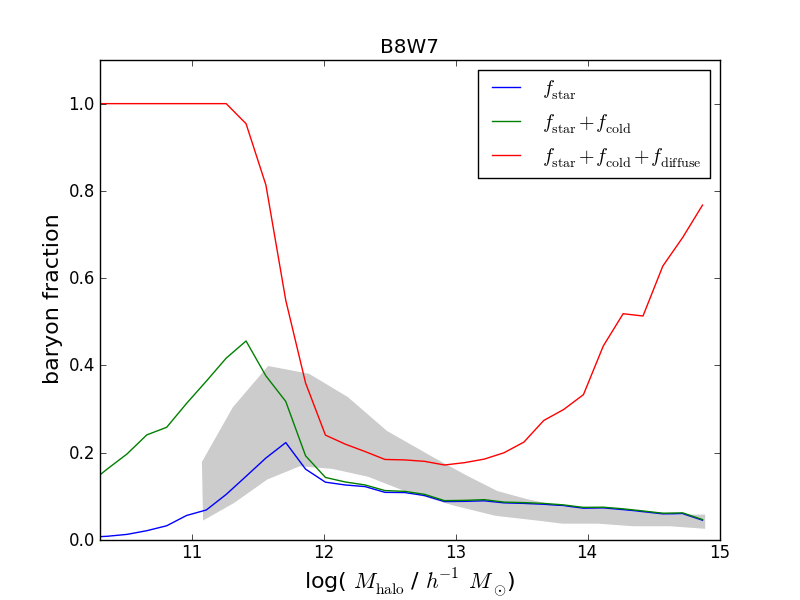}
\\
\includegraphics[scale=0.35,angle=0]{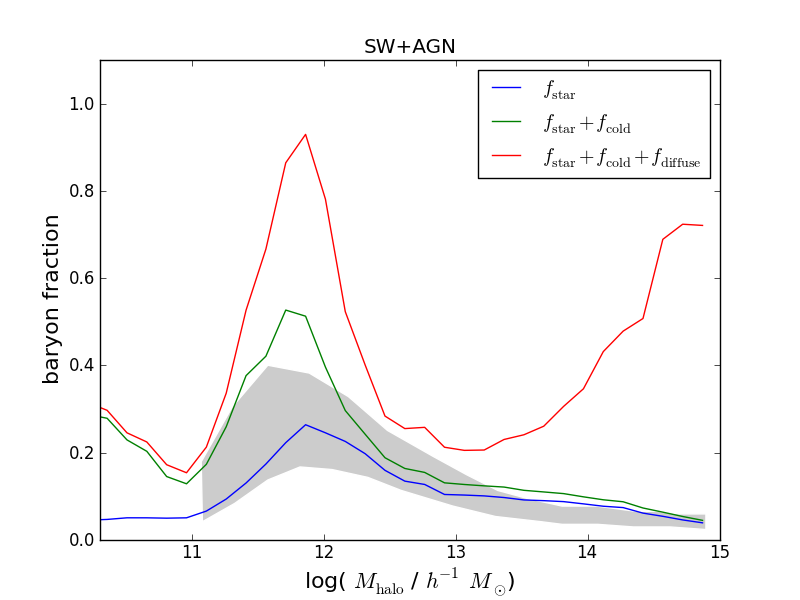}
\\
\includegraphics[scale=0.35,angle=0]{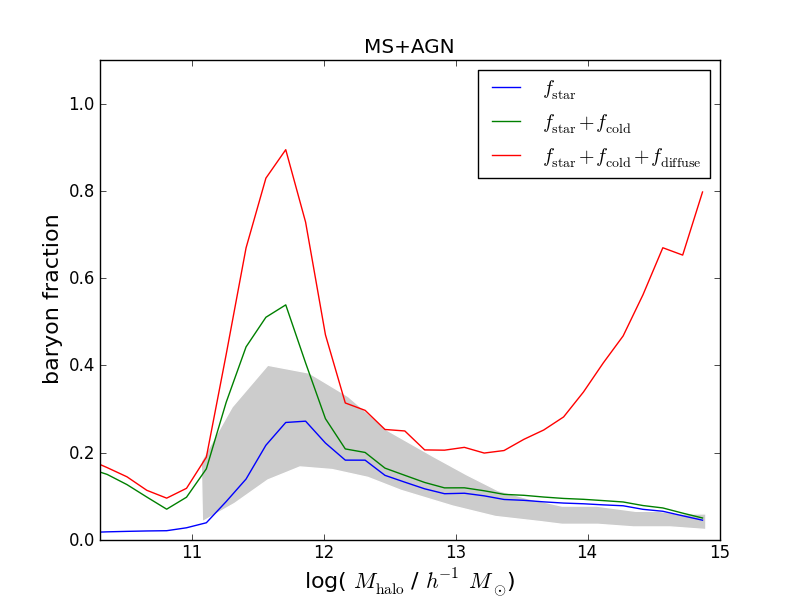}
\\
\end{tabular}
\end{center}
\caption{The fraction of baryons (relative to $\Mhalo{\Omega_b/\Omega_0}$)
in different phases as a function of halo mass. The lines are cumulative
showing, blue: stars, green: adding cold gas, red: adding hot and reheated gas. The three panels show the B8W7, SW
and MS models (all including the ``hot-halo'' mode of AGN feedback). The models include the effect of hot-halo AGN feedback, as shown in Fig.~\ref{fig:with_agn}.
The shaded region indicates the total stellar mass content of haloes measured in observational data-sets (see text for details).  
}
\label{fig:baryons}
\end{figure}

As we have seen, in the successful models, only a small fraction of baryons are locked up into stars. It is of interest to examine the phase of the remaining baryons in the model, and
this is illustrated in Fig.~\ref{fig:baryons}. We focus on the same three models that provided a good description of the SMF above $\Mstar \sim 10^9 \hMsol$: B8W7, SW $(\vw200,\b200) = (200,4)$ and MS $(\vw200,\b200) = (200,16)$. The SMFs of these models are shown in Fig.~\ref{fig:with_agn}. 

The distribution of baryonic matter between phases is plotted as a function of halo mass. We include all baryons associated with the halo by the model, and define the baryon fraction as the mass of baryons, $M_b$ divided by the cosmic average (ie., $ f_b = {M_b/ \Omega_b \Mhalo}$).  In the absence of feedback, the baryons collapse with the dark matter, and all haloes would be baryonically `closed' with fractions of order unity \citep{crain2007}.  The lines are cumulative so that the mass in stars is shown as a blue line.  The green line adds the contribution of cold gas, and the red line includes the contribution from diffuse halo gas (ie., `hot' plus `reheated' components in the terminology of Bow06)
This gives the total baryon mass associated with the halo.  Note that, because of the expulsion feedback considered in these models, the total mass of baryons in the halo need not reach the cosmic value.

As we should expect, given that the models reproduce the $\Mstar-\Mhalo$ relations deduced from sub-halo abundance matching, all the models have a similar peak in the fraction of baryons locked into stars around $\Mhalo \sim 10^{11.5}\hMsol$ (solid lines). In the B8W7 model, this declines rapidly to lower masses, and declines to an almost constant value ($\sim 0.08$) in higher mass systems.  In the SW model, the excess abundance of small galaxies means that the decline to lower mass haloes is less steep, reaching a constant value of 0.05 in haloes smaller than $10^{11}\hMsol$.  The MS model is intermediate between the two, as we should expect. The amplitude of the peak is smaller in the B8W7 case than in the other models. These differences are largely due to the contribution from very faint satellite galaxies ($\Mstar < 10^9 \hMsol$) in the SW and MS models, although it is also noticeable that the peak in stellar abundance is broader in these systems.  This is consistent with all of the models giving a similar match to the mass function because of the interaction between the halo abundance and the scatter in the $\Mstar$--$\Mhalo$ relation.

The most striking difference between the plots lies in the behaviour of
the total baryon contributions below $\Mhalo \sim 10^{11.5}\hMsol$. In
the B8W7 model (in this regime), all of the baryons reside in the host halo, and the diffuse gas makes up the majority of the baryons.  
It is questionable whether such a
large halo component (which we refer to as the `circum-galactic
medium', CGM) is compatible with observational data. Even though
recent observations from the Cosmic Origins Spectrograph (COS) on
\textit{HST} suggest that there is a significant CGM
\citep{tumlinson2011}, it is likely to account for ``only'' a smilar fraction to the mass in the galaxy's stars (ie., $\sim$10\% of $\Omega_b$, see \cite{prochaska2011}
for discussion). In contrast, the SW and MS models are able to eject much of the material from the halo so that low mass galaxies contain little diffuse gas in their haloes, and most of the baryons are locked into cold gas. As cold gas is ejected from these galaxies it escapes from the halo. Nevertheless, the feedback scheme in these models has been tuned to match the observed stellar mass function, so that the outflowing wind stalls as the halo mass approaches $10^{12}\hMsol$, resulting in a peak in the abundance of diffuse gas that matches the peak in the stellar fraction.
At masses above $\sim 10^{12}\hMsol$, material is driven out of the halo by AGN feedback in all three models, resulting in a second dip in the abundance of diffuse gas. As discussed in Bow08, this results in a good match to the X-ray luminosities of groups and clusters. The rapidly rising gas fractions accounting for the tilt in X-ray scaling relations compared to the `self similar' predictions.  

%{\bf ZZZ since we're not using Claudia's model here, the fraction in cold gas is a bit
%arbitrary, so I've not stressed it in the discussion.}

{\rgb In order to compare the models with observational data, we show measurements of the 
stellar mass fraction of haloes from \citet{leauthaud2011} as a shaded region in
Fig.~\ref{fig:baryons}. This analysis is based on combining a sub-halo
occupation distribution model and integrated stellar mass measurements
for a sample of COSMOS X-ray detected groups, for which halo masses have
been determined using a weak lensing analysis. In contrast to some
previous results (eg., \citet{giodini2009,gonzalez2007}), their careful
analysis shows that the stellar mass content of high mass haloes is
low and comparable to the mean stellar mass fraction of the universe
as a whole (as would be expected on quite general grounds \citep{balogh2008}).
The shaded region indicates the plausible systematic uncertainties in
the analysis. The survey includes only directly detected galaxies, and could
under-estimate the stellar mass content of some haloes with
significant intra-cluster light. However, the correction for
intra-cluster light is likely to be small and certainly less than 50\%
\citep{zibetti2005, mcgee2010}. As \citet{leauthaud2011}
discuss the differences from previous work arises from the treatment
of stellar populations and satellite galaxies, not from the
contribution of the intra-cluster light.

All of the models compare with this observational data reasonably
well. The SW and MS models fit particularly well, with a peak in the 
stellar mass fraction close to  $\sim 10^{12}\hMsol$. At higher
masses, the observations show a shallow decline, which is also seen
in the models. Differences between the model and the data could
certainly be accounted for by the missing intra-cluster light
distribution.  The B8W7 model show a somewhat sharper decline than is
evident in the data, together with a peak in the stellar mass fraction
that is shifted to lower masses. However, this part of the
observational region is dominated by the scatter in the $\Mstar$--$\Mhalo$ relation: all of the models presented reproduce the observed
SMF, as seen in Fig.~\ref{fig:with_agn}.
}

{\rgb
In summary, the different feedback schemes result in significant
differences in the distribution of mass between the phases 
of baryonic matter. In the B8W7 model, haloes below  $\sim 10^{11.5}\hMsol$
retain the cosmic baryon fraction, mostly in the form of a diffuse
gas halo. In the SW and MS models, most baryons have been ejected
from low mass haloes. In higher mass haloes, all three models suggest
that the stellar fraction should quickly converge to the average value.
The shape of the models fits well with observational measurements.
}

\subsection{Metal Enrichment of the Inter-Galactic Medium}

\begin{figure*}
\begin{center}
\begin{tabular}{cc}
\includegraphics[scale=0.35,angle=0]{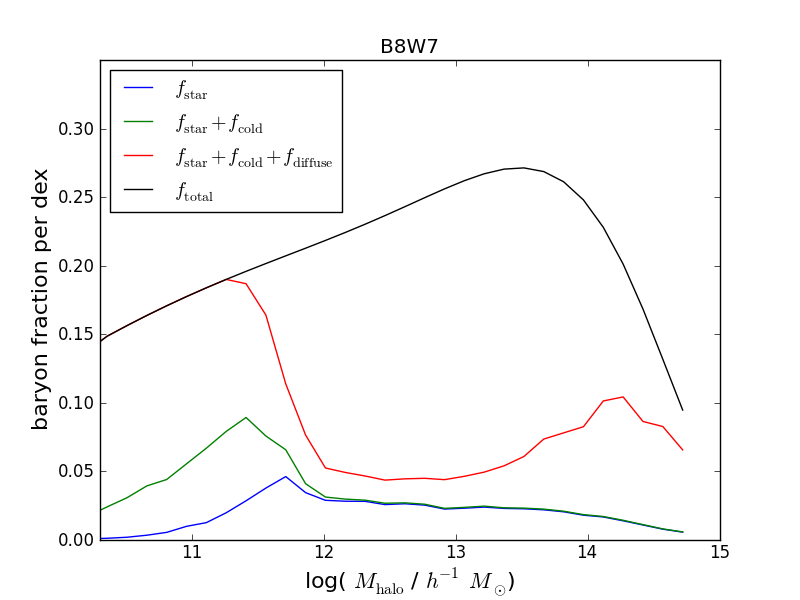} &
\includegraphics[scale=0.35,angle=0]{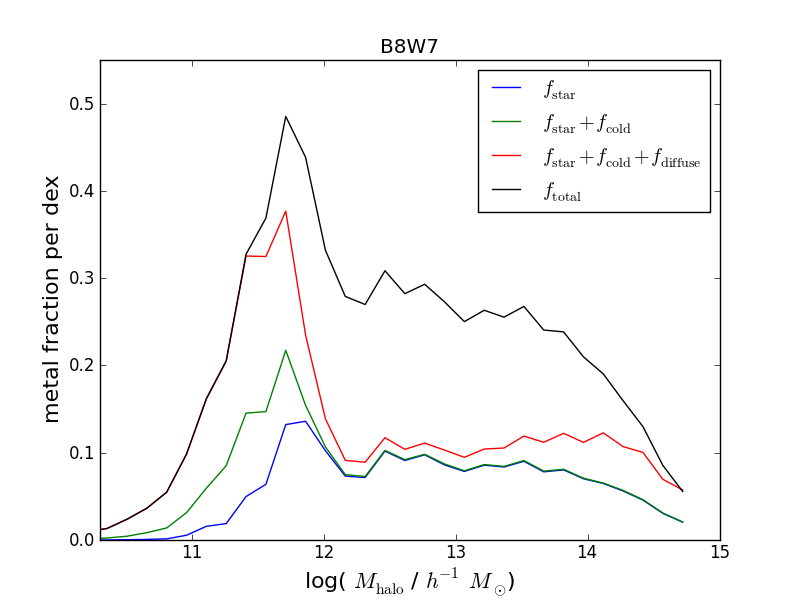}
\\
\includegraphics[scale=0.35,angle=0]{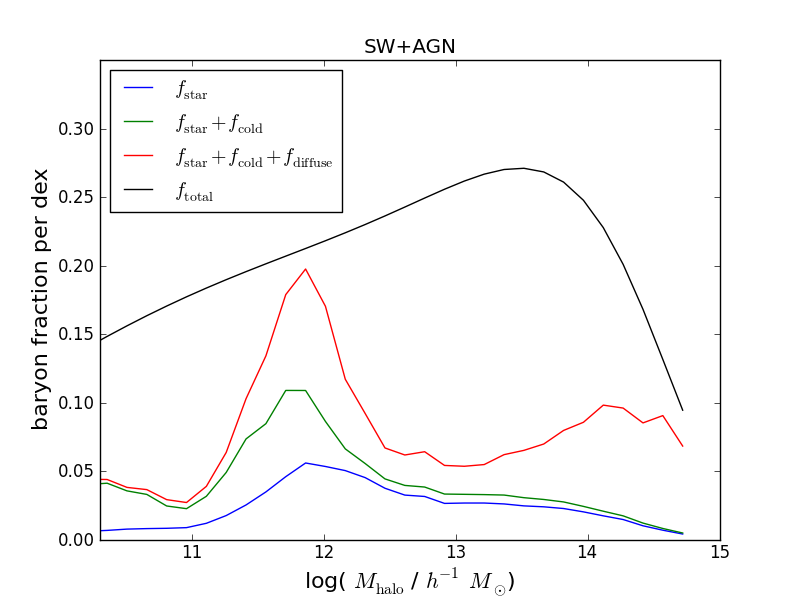} &
\includegraphics[scale=0.35,angle=0]{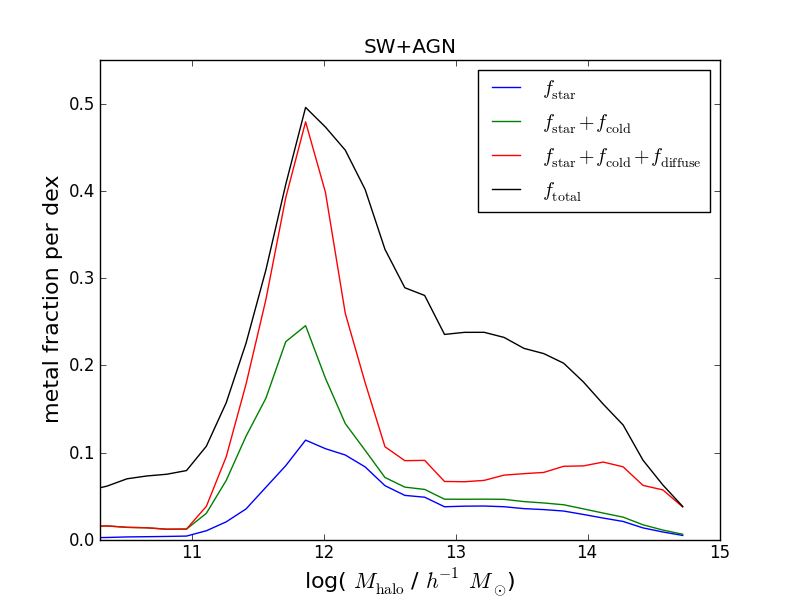}
\\
\includegraphics[scale=0.35,angle=0]{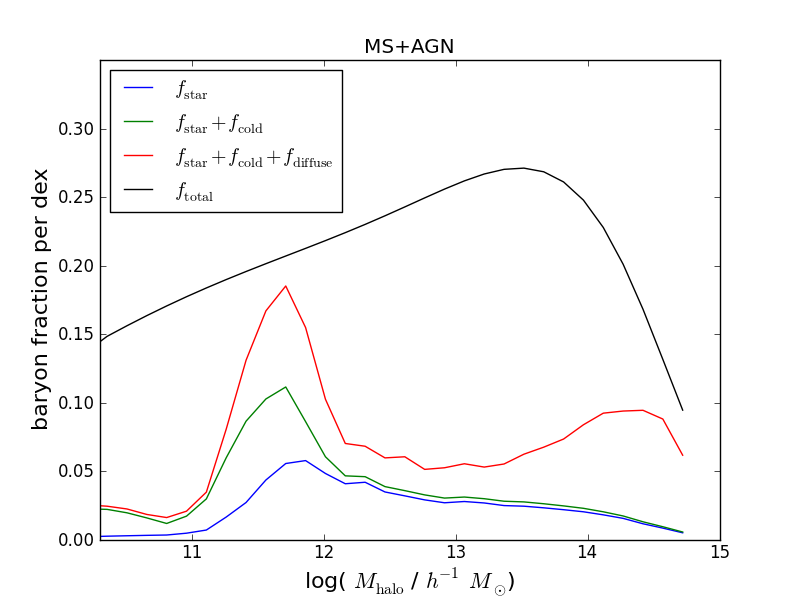} &
\includegraphics[scale=0.35,angle=0]{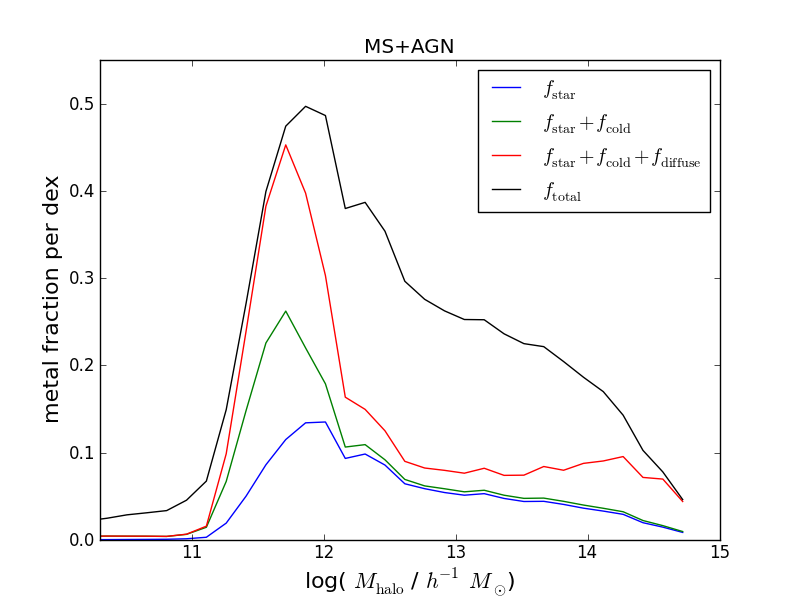}
\\
\end{tabular}
\end{center}
\caption{{\rgb (left-hand panels): The contribution of baryons  
in different phases to the total baryon content of the universe at $z=0$,
plotted as a function of halo mass. The lines are cumulative
showing, blue: stars, green: adding cold gas, red: adding hot and reheated gas,
and can be compared to the mass fraction (per halo mass bin) in different phases shown in
Fig.~\ref{fig:baryons}. The solid black line shows the total
contribution associated with haloes of a given mass in the absence of
any mass loss.  The difference between the red and black lines
indicates the contribution to the total intergalactic baryon mass that
has been expelled from haloes. (right-hand panels): the contribution
of haloes and phases to the total metal mass content of the universe
at $z=0$. The
black line indicates the metal mass that would be associated with
haloes in the absence of mass ejection, so that the difference between
the black and red lines indicates the contribution to intergalactic
metals. The figure shows that the contribution to intergalactic metals 
is dominated by the mass expelled from galaxy groups by AGN feedback
in all models. This dominates over the metals that are expelled from
the haloes of low mass galaxies.
}}
\label{fig:baryons2}
\end{figure*}

{\rgb Although the models present have all been developed in order to
explain the galaxy population of the universe and the X-ray scaling relations of galaxy clusters, the metal enrichment of the intergalactic medium could offer an important discriminator between the models. Indeed a major motivation
for considering powerful winds is to explain the wide-spread metal enrichment of the Universe. Recent observations
have shown that metals are widely distributed in the inter-galactic medium.  For example, \citet{prochaska2011} show that the low redshift metal absorption line cross-section
is compatible with 0.3 L* galaxies being surrounded by a metal enriched
halo extending out to $\sim300$~kpc (ie., well beyond the virial radius of
the galaxy). 
Quantitative comparison with these data is fraught with difficulties, however,
since the observations must be corrected for the ionisation state
of the absorbing clouds. This issue is further complicated by the 
multiphase nature of the absorbers.  The inherent correlation
of bright and faint galaxies must also be taken into account, particularly
if smaller galaxies dominate the enrichment. In addition, the phenomenological models include a number of simplifying assumptions that 
make quantitative comparison with the data difficult. In particular,
the models we have presented here assume that enrichment and metal
re-cycling is instantaneous and that metals are distributed uniformly
through the  galaxy's halo. These limitations must be carefully taken into
account in a detailed comparison, and we will not attempt this here.  
Nevertheless, it is instructive to briefly examine the differences
between one model and another.

%The contribution distribution of metals between different mass haloes
%is illustrated in Fig.~\ref{fig:metals}.
All of the models result in significant ejection of metals
from the galaxy disk, and there is little to distinguish between the 
models on the basis of the mass of metals ejected from the disk. 
Rather, the important discriminator is the distribution of the 
metals. In the B8W7 models the metals remain trapped in the 
dark matter potential of the galaxy, In the SW and MS models, metals
ejected from small galaxies will be more widely distributed. At face
value, the observed wide-spread distribution of metals would favour 
the SW or MS models.
}

{\rgb
In practice, however, the situation is more complex. At low redshift, 
in all of these models, most of the intergalactic metals are
associated with material that has been ejected from galaxy groups 
by the action of AGN feedback. 
The situation at low redshift is illustrated in Fig.~\ref{fig:baryons2}. The panels on the left
hand-side show the distribution of the baryon mass between haloes
of different mass, such that the integral under the curve gives the 
total baryonic mass of the universe in haloes more massive than
$10^{10}\hMsol$.
The coloured lines illustrate
the division of the baryons between different phases for the different
models. The blue, green and red lines show the cumulative contribution
from star, cold gas and diffuse gas, as previously discussed. 
The black line shows the contribution to the baryon mass  in the
absence of expulsion. 
It is noticeable that this is widely distributed,
with all haloes less massive than $10^{14}\hMsol$ making a similar
contribution to the total mass content.  Because the distribution
of mass between different haloes is rather flat, the coloured lines 
bear close resemblance to those in Fig.~\ref{fig:baryons}, and
attention can be focused on the difference between the red and black
lines. Since the scale is linear, this is the mass contribution from
material that has been ejected from haloes. In the B8W7 model, this
contribution only comes from the material expelled by AGN in haloes 
more massive than $10^{12}\hMsol$, while in the SW and MS models, 
expulsion feedback means that there is a second important contribution 
from haloes less massive than $10^{11.5}$.  The integral of the
difference between the red and black curves measures the total mass
fraction ejected from haloes.  We find that the fractions are 52\%,
61\% and 66\%, for the B8W7, SW and MS models respectively 
(for haloes more massive than $10^{10}\hMsol$).  

The panels on the right show the equivalent plot for the contribution
to the metal content of the Universe. The line colouring follows that
used in the left-hand panels. The black line
shows the contribution from stars that are now in haloes whose masses
are given on the horizontal axis. Of course, when the metals were 
produced, the stars may have been in lower mass haloes. The metals 
locked into stars, cold gas and diffuse gas are shown by the coloured
lines, and the area under the curve gives the total contribution.
In contrast to the left hand panels, the metal
contributions are much more strongly weighted to haloes more massive
than $10^{12}\hMsol$, which reflects the low efficiency of star
formation in smaller haloes, and thus the low rate of metal
production. As a result, the contribution to the global metal budgets from
low mass haloes is strongly down-weighted compared to the contribution
from galaxy-group haloes (driven by AGN feedback, see below). The fraction
of metals that have been lost from haloes is given by the difference
between the red and black lines. 
To be clear, the difference between the lines corresponds to the
fraction of metals that are absent from haloes of the mass given on
the x-axis at $z=0$.  Some of these metals may have been ejected from
projenitor haloes of smaller mass.
We find that the models have total
intergalactic metal fractions of 42\%, 45\% and  46\% for the B8W7, SW
and MS models respectively. In the B8W7, the black and red curves are
superposed below a halo mass of $10^{11.5}\hMsol$, but the results are
similar for all the models, with by far the largest contribution coming from
the group-scale haloes.  Above $10^{12}\hMsol$ the behaviour of the
models is similar since all the models share a common AGN feedback
scheme. In the absense of AGN feedback very few metals are ejected
from such haloes.  Focussing on
lower mass haloes, greater differences are apparent. In particular,
the SW and MS models have an extended tail making a useful contribution to the 
intergalactic metals. This tail is not present in the B8W7 model,
both because the material remains trapped in the halo potential and
because the SMF is much flatter in the B8W7 model
than in either the SW or MS models. Nevertheless, low mass haloes make
a relatively little contribution to the total metal content of the universe
in all the models.

\begin{figure}
\begin{center}
\begin{tabular}{c}
\includegraphics[scale=0.35,angle=0]{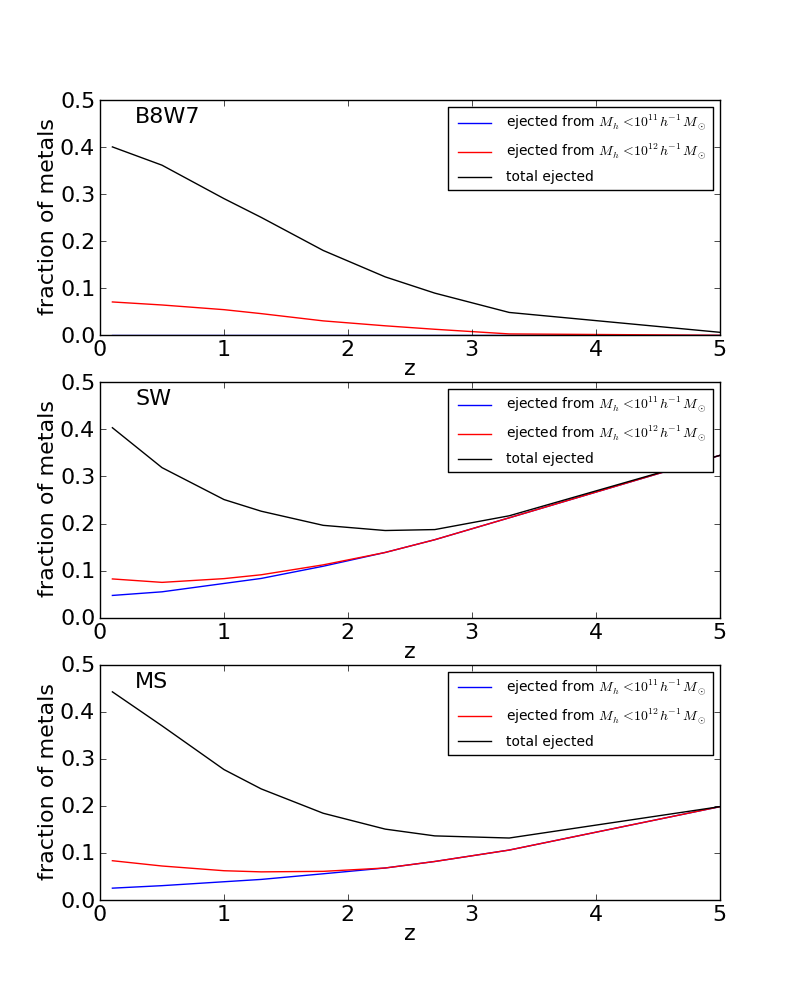}
\end{tabular}
\end{center}
\caption{The fraction of metals that have been expelled from haloes as a
  function of redshift for the three different models. The black line
  shows the total fraction of ejected metals, while the red and blue
  lines show the fraction ejected from haloes less massive than $10^{12} \hMsol$
  and $10^{11} \hMsol$ respectively. At low redshift the
  inter-galactic metal content is dominated by ejection from galaxy
  group and cluster haloes by the ``hot halo'' mode of feedback. At
  high redshift there are few such massive haloes this ejection
  mechanism is unimportant. As a result there are few intergalactic
  metals in the B8W7 model at $z>3$. In the SW and MS models, the
  fraction of metals ejected from haloes is much higher at this
  redshift, primarily as the result of ejection from dwarf galaxies.
}
\label{fig:igm_metals_evo}
\end{figure}

At higher redshifts, the fraction of intergalactic metals drops
rapidly in all the models, reflecting the lower abundance of high mass
haloes with effective ``hot mode'' feedback. At these redshifts, a larger
fraction of metals come from dwarf galaxies in low mass halos and this is 
reflected
in greater differences in the inter-galactic metal fraction between the
models. In Fig.~\ref{fig:igm_metals_evo}, we show the evolution of the
fraction of metals that have been ejected from galaxies. 
To be specific, we show the fraction of all metals
produced at higher redshifts, associating ejected metals with the halo in
which the stars that produced the metals are located at the specified
redshift. Metals produced by
high redshift galaxies may be recaptured by larger haloes at
low redshift, just as the original stars of the early galaxy may
have been mergered into a larger system. In principle, metals that are missing
from haloes of mass greater than $10^{12} \hMsol$ may have been ejected
at earlier time from a progenitor halo of much lower mass; however,
by re-running the models without AGN feedback, we have shown that this process
is negligible. In haloes more massive than $10^{12} \hMsol$, the loss
of metals from the halo is entirely due to the effect of AGN feedback.

The black line shows the total fraction of metals that are outside galaxy
haloes, while the coloured lines show how this fraction depends
on halo mass: the blue (red) line shows the material that has
been lost from haloes less massive than $10^{11} \hMsol$ ($10^{12}
\hMsol$) so that the contribution from
material ejected from small (large) galaxies can be seen. 
The blue line is not visible for the
B8W7 model because material is not expelled from galaxy haloes (the
small contribution from the green line is due to hot-halo mode
feedback in $\Mhalo \sim 10^{12}\hMsol$ systems). The area above the red line
shows the contribution from metals ejected by the ``hot halo'' mode of
feedback. Although this fraction declines quickly with increasing redshift it
dominates the galaxy contribution out to $z\sim2$, even in the MS and
SW models. At $z>3$, the IGM metal distribution reflects the ejection
of metals from dwarf galaxy haloes in these models, and the observed
widespread enrichment of the high redshift IGM \citep{aguirre2008,frank2010,steidel2011} seems to favour the expulsion 
models (SW and MS) over B8W7 \citep{bertone2005,booth2010,cen2011}.

Although these figures make for an interesting comparison
between models, quantitative comparison to the observed abundance of
intergalactic metals is extremely difficult. In particular, the current model
does not specify how far outside the 
group haloes the metal distribution will extend, and does not begin to
address its temperature and ionisation state. This issue can only
be satisfactorily addressed if we account for the non-spherical nature
of halo accretion and 
outflow.  Equally, the extended diffuse gas distribution
within haloes will also make a contribution to the metals seen in random
sight lines. It would therefore be premature to rule out any of the
models without careful consideration of the effects of galaxy clustering. This 
task is beyond the scope of the present paper but is clearly an
important avenue for future work.
}

{\rgb
In summary, all three models that we have considered eject the majority of metals
from the galaxy disk. The metals locked into stars and cold gas represent a 
relatively minor component of the universe. In the expulsion models, SW and MS,
the majority of metals completely escape from haloes smaller than $10^{11.2}\hMsol$. Thus these
metals are widely distributed across intergalactic space. In contrast, the B8W7 model
retains metals within lower mass haloes, creating a circumgalactic
medium. However, the dominant contribution to intergalactic metals in
all three models comes from the material ejected from group-scale
haloes by the action of AGN feedback. While current
observational data suggests that the metal haloes of galaxies extend well beyond their
virial radii (and thus that the haloes are more extended than the B8W7
model would suggest for individual galaxies), a more detailed study
of the effects of clustering (and the observed clustering of metal
lines) is required before the models can be distinguished on this basis.
}

\section{Conclusions}

`Feedback' is a fundamental component of galaxy formation models,
allowing us to understand the marked differences between the dark
matter halo mass function and the galaxy stellar mass function (SMF). 
In order to explain the observational data, galaxy winds
must strongly suppress the formation of low-mass
galaxies. Phenomenological (or ``semi-analytic'') models, such as GALFORM, have been shown to provide a good description of the galaxy mass function together with its evolution, the specific star formation rates of galaxies and the contribution of galaxies of different masses to the cosmic star formation rate density. In the standard GALFORM model, these successes are achieved by adopting a feedback parameterisation that varies strongly with system mass, such that the speed of the outflow tracks the mass of the halo while the mass loading of the wind decreases strongly with halo mass. In contrast most hydrodynamical models, including the GIMIC simulations, adopt a wind that is independent of the system mass.

In this paper, we have modified the semi-analytic code to implement feedback schemes similar to those usually adopted in hydrodynamic models. Our scheme allows the wind speed and mass loading to be adjusted and the consequent effect on the stellar mass function of galaxies, its evolution and other galaxy properties to be determined.  We focus on three particular models. Two models have fixed wind parameters model: the pGIMIC model has high wind speed and modest mass loading similar to the GIMIC hydrodynamical simulations \citep{crain2009}; the SW model has a slower wind speed (at the same mass loading).  We also consider a model (MS) in which the wind mass
loading scales with the inverse of the disk circular velocity. This is similar to the momentum scaling models of \citet{oppenheimer2008}, although we set the wind speed by keeping the total energy of the
wind (rather than its total momentum) fixed across halo masses. Initially, we also considered a model
with a $1/\vdisk^2$ dependence of mass loading (ES), but found this could not reproduce the observed stellar mass
function. We contrast these models with a model based on Bow06 and Bow08, but adapted to the WMAP7 cosmology used in this paper (B8W7). For each of these models we consider the effect of including a ``hot-halo'' mode of AGN feedback following the gas expulsion scheme of Bow08.

{\rgb The main results are as follows: 
\begin{itemize} 
\item We find that the phenomenological description of feedback that
  we use in our code reproduces many of the trends seen in 
  hydrodynamic simulations. This give us confidence that the
  phenomenological approach captures the key physical effects of
  galactic winds well and allows us to explore the parameter space of
  winds properties quickly and efficiently. Comparison of the
  phenomenological model and hydrodynamic simulation also allows us to
  better understand the key physical processes in galaxy formation. 
A model which
  uses a high wind speed and moderate mass loading (and does not
  include AGN feedback), reproduces the form of the SMF
  seen in the GIMIC numerical simulations well. Although
  the strong feedback suppresses the stellar masses of galaxies
  strongly, the effect is most prominent around $10^{10.5}\hMsol$, so
  that the resulting SMF has a dip in the abundance
  of galaxies at these masses. At lower masses, the
SMF rises steeply since star formation is too weakly
suppressed, while, at higher masses the wind is not sufficiently
energetic to escape the halo.  Using the phenomenological model, we
are able to explore how the dip in the mass function depends on the
assumed wind speed and mass loading.

\item The observed mass function is better matched by adopting a
relatively slow wind, as seen in the hydrodynamic simulations of
Op08. The SW model, with a wind speed of $180\kms$ and a wind mass loading
of 8, results in a good match to the observed galaxy mass
function over the range $10^9-10^{11} \hMsol$. Introducing the AGN
scheme developed in Bow08 suppresses the formation of galaxies from
cooling in hot gas haloes, resulting in a good match to the high-mass
turn-over of the mass function. The combined (AGN plus
supernova-driven wind) model provides a good description of galaxy 
abundance down to $10^9 \hMsol$ (but rises much more steeply
than the observational data at lower masses). Although this is
encouraging, we find that the model performs poorly compared to B8W7 in
several other aspects. In particular, when we
examine the specific star formation rates in this model, we see that
the fixed feedback scheme results in a kink in the specific-star
formation rates of galaxies at around a stellar mass of $10^{9.5}
\hMsol$, such that low mass galaxies have a factor $\sim 5$ lower star
formation rates than their massive counter-parts.  This is not seen in
observational data.  The model also tends to under represent the
abundance of massive ($\Mstar > 10^{11}\hMsol$) galaxies at $z>1$ compared to the
B8W7 model and the observational data.

\item We also consider a model in which the wind mass-loading scales with the inverse of the circular
velocity of the disk.  This model results in a more subtle transition
between the regime where material easily escapes the halo and that in
which it stalls. With suitable choice of parameters, and the inclusion
of the AGN feedback scheme, the model reproduces the observed galaxy
mass function above $\Mstar \sim 10^9 \hMsol$. However, although the
slope of the SSFR$-\Mstar$ relation is weaker
than that seen in the SW model, the trend is reversed compared
to the slightly rising relation seen in observational data. 
The model tends to under-predict the abundance of massive, $z>1$ galaxies 
compared to B8W7.  The deficit is, however, relatively small and
the MS model could be compatible with the observational data if the
random uncertainties in the stellar mass estimates are greater than
0.2~dex. 

\item Feedback from AGN may have two very distinct forms. The ``hot
  halo'' (or ``radio'') mode feedback suppresses the supply of gas
  from cooling haloes. This is a key component of the successful
  models we present. However, it is interesting to investigate if this
  can be replaced by a ``starburst'' or ``QSO'' mode of feedback. We
  implement this by enhancing the wind speed and mass loading during
  starburst events. We find that while this mode may enhance
  the feedback from quiescent star formation, it does not introduce a
  characteristic break in the galaxy mass function and cannot be seen
  as an alternative to the ``hot-halo'' mode feedback. 
  {\rgb Similar conclusions are reached
    by \citet{gabor2011}, when they introduce a comparable feedback
    scheme in hydrodynamical models.}

\item We have also investigated the star formation history of the each of the
  models, and its dependence on the stellar mass. Looking at the
  integrated star formation rate, the main difference lies in the
  strength of the decline between $z=2$ and the present day.  This is
  strongest in the B8W7 model. Nevertheless, given the uncertainties in the observational
  estimates, both the MS and B8W7 model seem to provide a reasonable
  description. However, the models show a much greater variety in
  behaviour when the results are broken down by stellar mass. For this
  reason, comparison of the total star formation rate may be miss-leading 
  since the observational results assume large extrapolations to
  account for galaxies that are not directly observed.  It is
  therefore cleaner to compare the contribution to the total star formation
  rate density at each epoch as a function of stellar mass. None of the models 
  provide a perfect description of the data. At low redshift,
  the distribution is broader in the B8W7 model than is the case for
  the SW or MS models. Observed contribution from low mass galaxies is
  better described by B8W7, but the model contains too few high star
  formation rate galaxies at low redshift. At $z\sim2$, the situation
  is reversed, with all the models showing a similar contribution from
  low mass galaxies, and the primary
  difference being the paucity of high stellar mass galaxies in the MS
  and SW models. 

\item The differences between the models are emphasised by comparing
  the mass fractions of baryons in different phases as a function of
  halo mass.  The conventional model, B8W7, assumes that haloes less
  massive than $10^{11.5} \hMsol$ are baryonically closed. Almost all
  of the baryons in low mass haloes are placed in cool clouds in the
  halo (because of the short halo cooling time). In contrast the SW
  and MS, expel much of their baryonic content from the halo. The
  fraction of baryons retained in the halo has a maximum of 80\% at
  $\Mhalo = 10^{11.7} \hMsol$ and drops rapidly to a minimum of $\sim
  10\%$ at $\Mhalo = 10^{11} \hMsol$. At the minimum, the majority of
  baryons are predominantly stars and cold gas. At masses above
  $10^{12}\hMsol$, the ``hot-halo'' mode of feedback takes over
  expelling the halo material, resulting in low baryon fractions as
  discussed in Bow08, and in good agreement with X-ray observations of
  groups and clusters. The halo baryon fractions of the MS and SW
  models offer a good explanation of the low abundance of HI clouds around M31
  \citep{sancisi2008}. Further work is required to compare the model
  predictions to such data and to the circumgalactic gas haloes inferred from
  absorption line systems at low redshift \citep{tumlinson2011} since
  the ionisation state of the extra-galactic gas must be carefully
  computed and combined with X-ray limits on the emission from
  galactic haloes \citep{crain2010}. In all of these models, the
  recycling of previously ejected gas plays an important role. 

% I tried to create an ejection model that had similar scaling to B8W7 - but the 
% problem is that the acceptable models eject little (or no) material from small 
% haloes. If this is turned up, the model cannot produce enough L* galaxies. 
% This is a consequence of assuming that the speed of the wind tracks the escape velocity 
% - to make ejection work, the model needs to have higher relative windspeed in small 
% haloes compared to large ones.  Try a model in which the wind speed has MS scaling, 
% but the mass loading increases like the ES model.

{\rgb
Another potential discriminant between the models is the distribution of metals in the 
ISM. All the models predict that a major fraction of the metals produced by stars
will reside outside galaxies. In the absence of AGN feedback, the B8W7
predicts that the metals will be confined within galaxy haloes, while it will be more wide-spread in the
expulsion models (SW and MS). We find, however, that at $z<2$ the major
contributor of extra-galactic metals is the AGN powered hot-halo feedback that expels
diffuse gas from galaxy groups.  This material is highly enriched
compared to the winds from low-mass galaxies: combined with the high fraction of the
total baryon budget that has been expelled from groups, this
component dominates the diffuse metal content of the low redshift
Universe. At $z>3$, AGN feedback makes little contribution to the
intergalactic fraction of metals and the observed wide-spread
distribution of metals at this epoch favours the expulsion feedback
schemes of the MS and SW models. However, the nature of our
phenomenological model makes it difficult to predict the dispersal of the
expelled material, and it is unclear how far from the parent galaxy groups the
metals will be spread \citep{booth2010}.  Careful consideration of the dynamics of the outflow is needed to make a meaningful comparison with observational data on metal absorption lines.}

\end{itemize}
}

{\rgb In summary, although we have introduced a feedback scheme that
  reproduces the results of hydrodynamical simulations well, we find
  that the original wind scheme of Bow06 produces a better match to
  observational data on the stellar mass content of haloes, the
  specific star formation rates of galaxies and the evolution of these
  quantities. 
Comparison of the models with the observational data highlights several
aspects of the observational data that are not consistent with any of
the models considered. Firstly, none of the models reproduce the rapid
drop in the normalisation of the observed SMF between $z=0$
and $z=1-2$. Secondly, although the B8W7 model reproduces the observed
dependence of the specific star formation rate on stellar mass better than the 
SW or MS models, none reproduce the observed tendency for lower mass
galaxies to have higher SSFR than their massive counter-parts. Both of
these trends suggest that the effective wind speed (relative to the
halo escape velocity) should increase with redshift. One possibility 
is that this might arise because of the more clumpy concentration of 
star forming regions in high redshift galaxies. This is an interesting
possibility for further investigation.
There is also
  no intrinsic reason why galaxy winds should be a simple power-law,
  and the observational suggest we should explore feedback schemes in
  which the exponent of the mass dependence varies with system.  
  For example, the problem of the SW and MS models over
  producing galaxies below $10^9\hMsol$ could be solved by introducing
  a strong mass dependence to the feedback below this
  limit. Rather than introducing phenomenological modifications to the 
  feedback scheme, an alternative approach would be to simulate idealised galaxies at high (10 pc), or
  ultra-high (0.1pc) resolution and to extract suitable parameterisations for the outflow
  \citet{hopkins2011,stringer2011b,creasey2011}.
}

{\rgb
In future, we will combine the expulsion mode of feedback considered
here with developments of the GALFORM code \citep{lagos2010} to more accurately trace the cold gas content of the universe. We will also make a one-to-one comparison of the formation histories of galaxies from GIMIC, and the forth-coming suite of EAGLE simulations, with the GALFORM code. Our aim is {\it not only} to develop the GALFORM model as an emulator of numerical simulations (eg., \citet{bower2010}) but also to use it as a fundamental tool for understanding the key components of successful galaxy formation models. Only by representing the evolution of galaxies as a simple set of coupled differential equations can we claim to have understood the galaxy formation problem, and to have separated the key processes from the details.
}

\section*{Acknowledgments}

We thank our collaborators who have helped to develop the GALFORM
project, Shaun Cole, Carlton Baugh, Cedric Lacey, Carlos Frenk,
Claudia Lagos and Nikos Fanidakis.   We thank Ivan Baldry for allowing us
access to pre-release data on the galaxy mass function from the GAMA
survey, and thank the anonymous referee for their thoughtful comments. 
AJB acknowledges the support of the Gordon \& Betty Moore Foundation;  
RAC is supported by the Australian Research Council via a Discovery
Project grant; and RGB thanks the STFC for support through the rolling grant scheme.


\begin{thebibliography}{99}

\bibitem[\protect\citeauthoryear{Aguirre et 
al.}{2008}]{aguirre2008} Aguirre A., Dow-Hygelund C., Schaye J., 
Theuns T., 2008, ApJ, 689, 851 

\bibitem[\protect\citeauthoryear{Balogh et al.}{2001}]{balogh2001} Balogh M.~L., Pearce F.~R., Bower R.~G., Kay S.~T., 2001, MNRAS, 326, 1228 

\bibitem[\protect\citeauthoryear{Balogh et al.}{2008}]{balogh2008} 
Balogh M.~L., McCarthy I.~G., Bower R.~G., Eke V.~R., 2008, MNRAS, 385, 
1003 


\bibitem[\protect\citeauthoryear{Baugh et al.}{2005}]{baugh2005} Baugh C.~M., Lacey C.~G., Frenk C.~S., Granato G.~L., Silva L., Bressan A., Benson A.~J., Cole S., 2005, MNRAS, 356, 1191 

\bibitem[\protect\citeauthoryear{Bell et al.}{2003}]{bell2003} Bell E.~F., McIntosh D.~H., Katz N., Weinberg M.~D., 2003, ApJS, 149, 289 

\bibitem[\protect\citeauthoryear{Benson et al.}{2006}]{benson2006} 
Benson A.~J., Sugiyama N., Nusser A., Lacey C.~G., 2006, MNRAS, 369, 1055 

\bibitem[\protect\citeauthoryear{Benson et al.}{2003}]{benson2003} Benson A.~J., Bower R.~G., Frenk C.~S., Lacey C.~G., Baugh C.~M., Cole S., 2003, ApJ, 599, 38 

\bibitem[\protect\citeauthoryear{Benson et al.}{2001}]{benson2001} 
Benson A.~J., Pearce F.~R., Frenk C.~S., Baugh C.~M., Jenkins A., 2001, 
MNRAS, 320, 261 

\bibitem[\protect\citeauthoryear{Bertone, Stoehr, 
\& White}{2005}]{bertone2005} Bertone S., Stoehr F., White S.~D.~M., 2005, MNRAS, 359, 1201 

\bibitem[\protect\citeauthoryear{Best et al.}{2007}]{best2007} 
Best P.~N., von der Linden A., Kauffmann G., Heckman T.~M., Kaiser C.~R., 
2007, MNRAS, 379, 894 

\bibitem[\protect\citeauthoryear{Booth \& Schaye}{2009}]{booth_schaye09} Booth C.~M., Schaye J., 2009, MNRAS, 398, 53 

\bibitem[\protect\citeauthoryear{Booth et al.}{2010}]{booth2010} 
Booth C.~M., Schaye J., Delgado J.~D., Dalla Vecchia C., 2010, arXiv, 
arXiv:1011.5502 

\bibitem[\protect\citeauthoryear{Bower et al.}{2006}]{bower2006} Bower R.~G., Benson A.~J., Malbon R., Helly J.~C., Frenk C.~S., Baugh C.~M., Cole S., Lacey C.~G., 2006, MNRAS, 370, 645 

\bibitem[\protect\citeauthoryear{Bower, McCarthy, \& Benson}{2008}]{bower2008} Bower R.~G., McCarthy I.~G., BensonA.~J., 2008, MNRAS, 390, 1399 

\bibitem[\protect\citeauthoryear{Bower et al.}{2010}]{bower2010} 
Bower R.~G., Vernon I., Goldstein M., Benson A.~J., Lacey C.~G., Baugh 
C.~M., Cole S., Frenk C.~S., 2010, MNRAS, 407, 2017 

\bibitem[\protect\citeauthoryear{Brinchmann et al.}{2004}]{brinchmann2004} Brinchmann J., Charlot S., WhiteS.~D.~M., Tremonti C., Kauffmann G., Heckman T., Brinkmann J., 2004, MNRAS, 351, 1151 

\bibitem[\protect\citeauthoryear{Brooks et al.}{2007}]{brooks2007} Brooks A.~M., Governato F., Booth C.~M., Willman B., Gardner J.~P., Wadsley J., Stinson G., Quinn T., 2007, ApJ, 655, L17 

\bibitem[\protect\citeauthoryear{Bundy, Ellis, \& Conselice}{2005}]{bundy2005} Bundy K., Ellis R.~S., Conselice C.~J., 2005, ApJ, 625, 621 

\bibitem[\protect\citeauthoryear{Cai et al.}{2009}]{cai2009} 
Cai Y.-C., Angulo R.~E., Baugh C.~M., Cole S., Frenk C.~S., Jenkins A., 
2009, MNRAS, 395, 1185 

\bibitem[\protect\citeauthoryear{Cen 
\& Chisari}{2011}]{cen2011} Cen R., Chisari N.~E., 2011, ApJ, 731, 11 

\bibitem[\protect\citeauthoryear{Chen et al.}{2010}]{chan2010} 
Chen Y.-M., Tremonti C.~A., Heckman T.~M., Kauffmann G., Weiner B.~J., 
Brinchmann J., Wang J., 2010, AJ, 140, 445 

\bibitem[\protect\citeauthoryear{Cirasuolo et 
al.}{2010}]{cirasuolo2010} Cirasuolo M., McLure R.~J., Dunlop J.~S., 
Almaini O., Foucaud S., Simpson C., 2010, MNRAS, 401, 1166 

\bibitem[\protect\citeauthoryear{Cole et al.}{2000}]{cole2000} Cole S., Lacey C.~G., Baugh C.~M., Frenk C.~S., 2000, MNRAS, 319, 168 

\bibitem[\protect\citeauthoryear{Crain et al.}{2007}]{crain2007} Crain R.~A., Eke V.~R., Frenk C.~S., Jenkins A., McCarthy I.~G., Navarro J.~F., Pearce F.~R., 2007, MNRAS, 377, 41 

\bibitem[\protect\citeauthoryear{Crain et al.}{2009}]{crain2009} Crain R.~A., et al., 2009, MNRAS, 399, 1773 

\bibitem[\protect\citeauthoryear{Crain et al.}{2010}]{crain2010} Crain R.~A., McCarthy I.~G., Frenk C.~S., Theuns T., Schaye J., 2010, MNRAS, 407, 1403 

\bibitem[\protect\citeauthoryear{Creasey et al.}{2011}]{creasey2011} Creasey P. et al., 2011, in prep.

\bibitem[\protect\citeauthoryear{Croton et al.}{2006}]{croton2006} Croton D.~J., et al., 2006, MNRAS, 365, 11 

\bibitem[\protect\citeauthoryear{Cucciati et 
al.}{2011}]{cucciati2011} Cucciati O., et al., 2011, arXiv, 
arXiv:1109.1005 

\bibitem[\protect\citeauthoryear{Dalla Vecchia \& Schaye}{2008}]{dv_schaye08} Dalla Vecchia C., Schaye J., 2008, MNRAS, 387, 1431 

\bibitem[\protect\citeauthoryear{Dav{\'e}, Oppenheimer, \& Finlator}{2011}]{dave2011} Dav{\'e} R., Oppenheimer B.~D., Finlator K., 2011, MNRAS, 867 

%ZZZ extra dave reference

%\bibitem[\protect\citeauthoryear{de Avillez %\& Breitschwerdt}{2004}]{deavillez2004} de Avillez M.~A., Breitschwerdt D., 2004, A\&A, 425, 899 

\bibitem[\protect\citeauthoryear{de Avillez \& Breitschwerdt}{2007}]{deavillez2007} de Avillez M.~A., Breitschwerdt D., 2007, ApJ, 665, L35 

\bibitem[\protect\citeauthoryear{Deason et al.}{2011}]{deason2011} 
Deason A.~J., et al., 2011, arXiv, arXiv:1101.0816 

\bibitem[\protect\citeauthoryear{Dekel 
\& Silk}{1986}]{dekel_silk86} Dekel A., Silk J., 1986, ApJ, 303, 39 

\bibitem[\protect\citeauthoryear{De Lucia et al.}{2006}]{delucia2006} De Lucia G., Springel V., White S.~D.~M., Croton D., Kauffmann G., 2006, MNRAS, 366, 499 

\bibitem[\protect\citeauthoryear{De Lucia et al.}{2010}]{delucia2010} De Lucia G., Boylan-Kolchin M., Benson A.~J., Fontanot F., Monaco P., 2010, MNRAS, 406, 1533 

\bibitem[\protect\citeauthoryear{Drory et al.}{2005}]{drory2005} Drory N., Salvato M., Gabasch A., Bender R., Hopp U., Feulner G., Pannella M., 2005, ApJ, 619, L131 

\bibitem[\protect\citeauthoryear{Efstathiou}{2000}]{efstathiou2000} Efstathiou G., 2000, MNRAS, 317, 697 

\bibitem[\protect\citeauthoryear{Fanidakis et al.}{2011}]{fanidakis2011} Fanidakis N., Baugh C.~M., Benson A.~J., Bower R.~G., Cole S., Done C., Frenk C.~S., 2011, MNRAS, 410, 53 

\bibitem[\protect\citeauthoryear{Font et al.}{2008}]{font2008} 
Font A.~S., et al., 2008, MNRAS, 389, 1619 

\bibitem[\protect\citeauthoryear{Font et al.}{2011}]{font2011a} Font A.~S., et al., 2011, arXiv, arXiv:1103.0024 

\bibitem[\protect\citeauthoryear{Font et al.}{2011}]{font2011} Font A.~S., McCarthy I.~G., Crain R.~A., Theuns T., Schaye J., Wiersma R.~P.~C., Dalla Vecchia C., 2011, arXiv, arXiv:1102.2526 

\bibitem[\protect\citeauthoryear{Frank et al.}{2010}]{frank2010} 
Frank S., Mathur S., Pieri M., York D.~G., 2010, AJ, 140, 835 

\bibitem[\protect\citeauthoryear{Gabor et al.}{2011}]{gabor2011} 
Gabor J.~M., Dav{\'e} R., Oppenheimer B.~D., Finlator K., 2011, MNRAS, 417, 
2676 

\bibitem[\protect\citeauthoryear{Genzel et al.}{2011}]{genzel2011} 
Genzel R., et al., 2011, ApJ, 733, 101 

\bibitem[\protect\citeauthoryear{Gilbank et al.}{2011}]{gilbank2011} Gilbank D.~G., et al., 2011, MNRAS, 414, 304 

\bibitem[\protect\citeauthoryear{Giodini et 
al.}{2009}]{giodini2009} Giodini S., et al., 2009, ApJ, 703, 982 

\bibitem[\protect\citeauthoryear{Gonzalez, Zaritsky, 
\& Zabludoff}{2007}]{gonzalez2007} Gonzalez A.~H., Zaritsky D., Zabludoff A.~I., 2007, ApJ, 666, 147 

\bibitem[\protect\citeauthoryear{Guo et al.}{2010}]{guo2010} Guo Q., White S., Li C., Boylan-Kolchin M., 2010, MNRAS, 404, 1111 

\bibitem[\protect\citeauthoryear{Guo et al.}{2011}]{guo2011} Guo Q., et al., 2011, MNRAS, 413, 101 

\bibitem[\protect\citeauthoryear{Hatton et al.}{2003}]{hatton2003} Hatton S., Devriendt J.~E.~G., Ninin S., Bouchet F.~R., Guiderdoni B., Vibert D., 2003, MNRAS, 343, 75 

\bibitem[\protect\citeauthoryear{Helly et al.}{2003}]{helly2003} Helly J.~C., Cole S., Frenk C.~S., Baugh C.~M., Benson A., Lacey C., Pearce F.~R., 2003, MNRAS, 338, 913 

\bibitem[\protect\citeauthoryear{Hopkins}{2004}]{hopkins2004} 
Hopkins A.~M., 2004, ApJ, 615, 209 

%\bibitem[\protect\citeauthoryear{Hopkins et 
%al.}{2008}]{2008ApJS..175..390H} Hopkins P.~F., Cox T.~J., Kere{\v s} D., 
%Hernquist L., 2008, ApJS, 175, 390 

\bibitem[\protect\citeauthoryear{Hopkins et 
al.}{2006}]{hopkins_hernquist2006} Hopkins P.~F., Hernquist L., Cox T.~J., 
Robertson B., Springel V., 2006, ApJS, 163, 50 

\bibitem[\protect\citeauthoryear{Hopkins, Quataert, 
\& Murray}{2011}]{hopkins2011} Hopkins P.~F., Quataert E., Murray N., 2011, arXiv, arXiv:1110.4638 

\bibitem[\protect\citeauthoryear{Hopkins 
\& Elvis}{2010}]{hopkins2010} Hopkins P.~F., Elvis M., 2010, MNRAS, 401, 7 

\bibitem[\protect\citeauthoryear{Jaacks et al.}{2011}]{jaacks2011} 
Jaacks J., Choi J.-H., Nagamine K., Thompson R., Varghese S., 2011, arXiv, 
arXiv:1104.2345 

\bibitem[\protect\citeauthoryear{Jones et al.}{2010}]{jones2010} 
Jones T.~A., Swinbank A.~M., Ellis R.~S., Richard J., Stark D.~P., 2010, 
MNRAS, 404, 1247 

\bibitem[\protect\citeauthoryear{Juneau et al.}{2005}]{juneau2005} Juneau S., et al., 2005, ApJ, 619, L135 

\bibitem[\protect\citeauthoryear{Karim et al.}{2011}]{karim2011} 
Karim A., et al., 2011, ApJ, 730, 61 

\bibitem[\protect\citeauthoryear{Katz, Weinberg, \& Hernquist}{1996}]{katz1996} Katz N., Weinberg D.~H., Hernquist L., 1996, ApJS, 105, 19 

\bibitem[\protect\citeauthoryear{Kauffmann, White, \& Guiderdoni}{1993}]{kauffmann1993} Kauffmann G., White S.~D.~M., Guiderdoni B., 1993, MNRAS, 264, 201 

\bibitem[\protect\citeauthoryear{Kere{\v s} et al.}{2009}]{keres2009} Kere{\v s} D., Katz N., Dav{\'e} R., Fardal M., Weinberg D.~H., 2009, MNRAS, 396, 2332 

\bibitem[\protect\citeauthoryear{Komatsu et al.}{2011}]{komatsu2011} Komatsu E., et al., 2011, ApJS, 192, 18 

\bibitem[\protect\citeauthoryear{Lagos et al.}{2010}]{lagos2010} Lagos C.~d.~P., Lacey C.~G., Baugh C.~M., Bower R.~G., Benson A.~J., 2010, arXiv, arXiv:1011.5506 

\bibitem[\protect\citeauthoryear{Leauthaud et 
al.}{2011}]{leauthaud2011} Leauthaud A., et al., 2011, arXiv, 
arXiv:1109.0010 

\bibitem[\protect\citeauthoryear{Li \& White}{2009}]{li_white09} Li C., White S.~D.~M., 2009, MNRAS, 398, 2177 

\bibitem[\protect\citeauthoryear{Marchesini et al.}{2009}]{marchesini2009} Marchesini D., van Dokkum P.~G., F{\"o}rster Schreiber N.~M., Franx M., Labb{\'e} I., Wuyts S., 2009, ApJ, 701, 1765 

\bibitem[\protect\citeauthoryear{Martin}{2005}]{martin2005} Martin 
C.~L., 2005, ApJ, 621, 227 

\bibitem[\protect\citeauthoryear{McGee 
\& Balogh}{2010}]{mcgee2010} McGee S.~L., Balogh M.~L., 2010, MNRAS, 403, L79 

\bibitem[\protect\citeauthoryear{McGee et al.}{2011}]{mcgee2011} McGee S.~L., Balogh M.~L., Wilman D.~J., Bower R.~G., Mulchaey J.~S., Parker L.~C., Oemler A., 2011, MNRAS, 413, 996 

\bibitem[\protect\citeauthoryear{McKee \& Ostriker}{1977}]{mckee_ostriker77} McKee C.~F., Ostriker J.~P., 1977, ApJ, 218, 148 

\bibitem[\protect\citeauthoryear{Moster et al.}{2010}]{moster2010} Moster B.~P., Somerville R.~S., Maulbetsch C., van den Bosch F.~C., Macci{\`o} A.~V., Naab T., Oser L., 2010, ApJ, 710, 903 

\bibitem[\protect\citeauthoryear{Mortlock et al.}{2011}]{mortlock2011} Mortlock A., Conselice C.~J., Bluck A.~F.~L., Bauer A.~E., Gr{\"u}tzbauch R., Buitrago F., Ownsworth J., 2011, MNRAS, 413, 2845 

\bibitem[\protect\citeauthoryear{Murray, Quataert, \& Thompson}{2005}]{murray2005} Murray N., Quataert E., Thompson T.~A., 2005, ApJ, 618, 569 

\bibitem[\protect\citeauthoryear{Oppenheimer 
\& Dav{\'e}}{2006}]{oppenheimer2006} Oppenheimer B.~D., Dav{\'e} R., 2006, MNRAS, 373, 1265 

\bibitem[\protect\citeauthoryear{Oppenheimer \& Dav{\'e}}{2008}]{oppenheimer2008} Oppenheimer B.~D., Dav{\'e} R., 2008, MNRAS, 387, 577 

\bibitem[\protect\citeauthoryear{Oppenheimer et al.}{2010}]{oppenheimer2010} Oppenheimer B.~D., Dav{\'e} R., Kere{\v s} D., Fardal M., Katz N., Kollmeier J.~A., Weinberg D.~H., 2010, MNRAS, 406, 2325 

\bibitem[\protect\citeauthoryear{Okamoto, Gao, \& Theuns}{2008}]{okamoto2008} Okamoto T., Gao L., Theuns T., 2008, MNRAS, 390, 920 

\bibitem[\protect\citeauthoryear{Okamoto, Nemmen, \& Bower}{2008}]{okamoto_nemmen2008} Okamoto T., Nemmen R.~S., Bower R.~G., 2008, MNRAS, 385, 161 

\bibitem[\protect\citeauthoryear{Pozzetti et al.}{2010}]{pozzetti2010} Pozzetti L., et al., 2010, A\&A, 523, A13 


\bibitem[\protect\citeauthoryear{Prochaska et al.}{2011}]{prochaska2011} Prochaska J.~X., Weiner B., Chen H.~-., Mulchaey J.~S., Cooksey K.~L., 2011, arXiv, arXiv:1103.1891 

\bibitem[\protect\citeauthoryear{Reddy 
\& Steidel}{2009}]{reddy2009} Reddy N.~A., Steidel C.~C., 2009, ApJ, 692, 778 

\bibitem[\protect\citeauthoryear{Rodighiero et 
al.}{2010}]{rodighiero2010} Rodighiero G., et al., 2010, A\&A, 518, L25 

\bibitem[\protect\citeauthoryear{Rubin et al.}{2011}]{rubin2011} 
Rubin K.~H.~R., Prochaska J.~X., M{\'e}nard B., Murray N., Kasen D., Koo 
D.~C., Phillips A.~C., 2011, ApJ, 728, 55 

\bibitem[\protect\citeauthoryear{Sancisi et al.}{2008}]{sancisi2008} Sancisi R., Fraternali F., Oosterloo T., van der Hulst T., 2008, A\&ARv, 15, 189 

\bibitem[\protect\citeauthoryear{Schaye et al.}{2010}]{schaye2010} Schaye J., et al., 2010, MNRAS, 402, 1536 

\bibitem[\protect\citeauthoryear{Schaye \& Dalla Vecchia}{2008}]{schaye_dv08} Schaye J., Dalla Vecchia C., 2008, MNRAS, 383, 1210 

\bibitem[\protect\citeauthoryear{Somerville et al.}{2008}]{somerville2008} Somerville R.~S., Hopkins P.~F., Cox T.~J., Robertson B.~E., Hernquist L., 2008, MNRAS, 391, 481 

\bibitem[\protect\citeauthoryear{Springel \& Hernquist}{2003}]{springel_hernquist03} Springel V., Hernquist L., 2003, MNRAS, 339, 289 

%\bibitem[\protect\citeauthoryear{Springel}{2010}]{springel_review} Springel V., 2010, ARA\&A, 48, 391 

\bibitem[\protect\citeauthoryear{Springel, Di Matteo, \& Hernquist}{2005}]{springel2005} Springel V., Di Matteo T., Hernquist L., 2005, MNRAS, 361, 776 

\bibitem[\protect\citeauthoryear{Springel, Frenk, \& White}{2006}]{springel2006} Springel V., Frenk C.~S., White S.~D.~M., 2006, Natur, 440, 1137 

\bibitem[\protect\citeauthoryear{Steidel et 
al.}{2011}]{steidel2011} Steidel C.~C., Bogosavljevi{\'c} M., 
Shapley A.~E., Kollmeier J.~A., Reddy N.~A., Erb D.~K., Pettini M., 2011, 
ApJ, 736, 160 

\bibitem[\protect\citeauthoryear{Stringer et 
al.}{2011}]{stringer2011b} Stringer M.~J., Bower R.~G., Cole S., 
Frenk C.~S., Theuns T., 2011, arXiv, arXiv:1111.2529 

%\bibitem[\protect\citeauthoryear{Stringer et 
%al.}{2011}]{stringer2011a} Stringer M., Cole S., Frenk C.~S., Stark 
%D.~P., 2011, MNRAS, 414, 1927 

\bibitem[\protect\citeauthoryear{van de Voort et 
al.}{2011}]{vandevoort2011} van de Voort F., Schaye J., Booth C.~M., 
Dalla Vecchia C., 2011, MNRAS, 415, 2782 

%\bibitem[\protect\citeauthoryear{van de Voort et 
%al.}{2011}]{vd_voort2011} van de Voort F., Schaye J., Altay G., 
%Theuns T., 2011, arXiv, arXiv:1109.5700 

\bibitem[\protect\citeauthoryear{Tumlinson et 
al.}{2011}]{tumlinson2011} Tumlinson J., et al., 2011, arXiv, 
arXiv:1111.3980 


\bibitem[\protect\citeauthoryear{White \& Frenk}{1991}]{white_frenk91} White S.~D.~M., Frenk C.~S., 1991, ApJ, 379, 52 

\bibitem[\protect\citeauthoryear{Weiner et al.}{2009}]{weiner2009} 
Weiner B.~J., et al., 2009, ApJ, 692, 187 

\bibitem[\protect\citeauthoryear{Weinmann, Neistein, 
\& Dekel}{2011}]{weinmann2011} Weinmann S.~M., Neistein E., Dekel A., 2011, MNRAS, 417, 2737 

\bibitem[\protect\citeauthoryear{Wiersma et 
al.}{2009}]{wiersma2009} Wiersma R.~P.~C., Schaye J., Theuns T., 
Dalla Vecchia C., Tornatore L., 2009, MNRAS, 399, 574 

\bibitem[\protect\citeauthoryear{Wiersma, Schaye, 
\& Smith}{2009}]{wiersma_schaye2009} Wiersma R.~P.~C., Schaye J., Smith B.~D., 2009, MNRAS, 393, 99 

\bibitem[\protect\citeauthoryear{Zibetti et 
al.}{2005}]{zibetti2005} Zibetti S., White S.~D.~M., Schneider 
D.~P., Brinkmann J., 2005, MNRAS, 358, 949 


\end{thebibliography}
\end{document}